\newcommand{\xBc}{\langle}
\newcommand{\xBe}{\rangle}

\newcommand{\xbP}{\Pi}

\newcommand{\xbS}{\Sigma}

\newcommand{\xba}{\alpha}
\newcommand{\xbb}{\beta}

\newcommand{\xbe}{\in}
\newcommand{\xbf}{\phi}

\newcommand{\xbm}{\mu}

\newcommand{\xbo}{\omega}

\newcommand{\xbq}{\psi}
\newcommand{\xbr}{\rho}
\newcommand{\xbs}{\sigma}
\newcommand{\xbt}{\tau}

\newcommand{\xCK}{\times}

\newcommand{\xCN}{\neg}
\newcommand{\xCO}{ }
\newcommand{\xCQ}{\emptyset}

\newcommand{\xCf}{\hspace{0.1em}}

\newcommand{\xcA}{\forall}

\newcommand{\xcC}{\not\subseteq}

\newcommand{\xcE}{\exists}

\newcommand{\xcM}{\not\models}

\newcommand{\xcP}{\not\rightarrow}

\newcommand{\xcS}{\bigcap}

\newcommand{\xcU}{\bigwedge}
\newcommand{\xcV}{\bigcup}

\newcommand{\xcb}{\subset}
\newcommand{\xcc}{\subseteq}

\newcommand{\xce}{\not\in}

\newcommand{\xcg}{\geq}
\newcommand{\xch}{\Rightarrow}
\newcommand{\xci}{\Leftarrow}

\newcommand{\xck}{\leq}

\newcommand{\xcm}{\models}
\newcommand{\xcn}{\hspace{0.2em}\sim\hspace{-0.9em}\mid\hspace{0.58em}}

\newcommand{\xco}{\vee}
\newcommand{\xcp}{\rightarrow}

\newcommand{\xcr}{\leftrightarrow}
\newcommand{\xcs}{\cap}
\newcommand{\xcu}{\wedge}
\newcommand{\xcv}{\cup}

\newcommand{\xcz}{\Box}

\newcommand{\xDB}{\\[2ex]}

\newcommand{\xDH}{\item }

\newcommand{\xdR}{\Re}

\newcommand{\xda}{{\cal A}}

\newcommand{\xdf}{{\cal F}}

\newcommand{\xdn}{{\cal N}}

\newcommand{\xdp}{{\cal P}}

\newcommand{\xdx}{{\cal X}}
\newcommand{\xdy}{{\cal Y}}
\newcommand{\xdz}{{\cal Z}}

\newcommand{\xEH}{ & }
\newcommand{\xEI}{\begin{itemize}}
\newcommand{\xEJ}{\end{itemize}}
\newcommand{\xEP}{ \\ }

\newcommand{\xEd}{\neq}

\newcommand{\xEh}{\begin{enumerate}}

\newcommand{\xEj}{\end{enumerate}}

\newcommand{\xeB}{\not\prec}

\newcommand{\xeb}{\prec}
\newcommand{\xec}{\preceq}
\newcommand{\xed}{\succeq}
\newcommand{\xee}{\succ}

\newcommand{\xex}{\upharpoonright}

\newcommand{\xfA}{\mid}

\newcommand{\xfI}{\mbox{I}}

\newcommand{\Xl}{\ldots}

\newcommand{\bl}{\begin{lemma} \rm}
\newcommand{\el}{\end{lemma}}
\newcommand{\br}{\begin{remark} \rm}
\newcommand{\er}{\end{remark}}
\newcommand{\be}{\begin{example} \rm}
\newcommand{\ee}{\end{example}}
\newcommand{\bco}{\begin{corollary} \rm}
\newcommand{\eco}{\end{corollary}}
\newcommand{\bc}{\begin{claim} \rm}
\newcommand{\ec}{\end{claim}}
\newcommand{\bfa}{\begin{fact} \rm}
\newcommand{\efa}{\end{fact}}
\newcommand{\bp}{\begin{proposition} \rm}
\newcommand{\ep}{\end{proposition}}
\newcommand{\bd}{\begin{definition} \rm}
\newcommand{\ed}{\end{definition}}
\newcommand{\bcs}{\begin{construction} \rm}
\newcommand{\ecs}{\end{construction}}
\newcommand{\bcd}{\begin{condition} \rm}
\newcommand{\ecd}{\end{condition}}
\newcommand{\bt}{\begin{theorem} \rm}
\newcommand{\et}{\end{theorem}}
\newcommand{\bn}{\begin{notation} \rm}
\newcommand{\en}{\end{notation}}
\newcommand{\bfi}{\begin{bild} \rm}
\newcommand{\efi}{\end{bild}}
\newcommand{\bsta}{\begin{statement} \rm}
\newcommand{\esta}{\end{statement}}
\newcommand{\bcom}{\begin{comment} \rm}
\newcommand{\ecom}{\end{comment}}
\newcommand{\bdia}{\begin{diagram} \rm}
\newcommand{\edia}{\end{diagram}}

\newcommand{\bfc}{\begin{figure}[htb] \begin{center}}
\newcommand{\efc}{\end{center} \end{figure}}

\makeindex

\documentclass{book}

\usepackage{amssymb,latexsym,epic,eepic,rotating}

\title{
Equilibria und weiteres Heiteres - Sept. 2011
}

\author{Dov M. Gabbay
\thanks{
Dov.Gabbay@kcl.ac.uk, www.dcs.kcl.ac.uk/staff/dg
} \\
King's College, London
\thanks{
Department of Computer Science, King's College London, Strand,
London WC2R 2LS, UK
} \\
and \\
Bar-Ilan University, Israel
\thanks{
Department of Computer Science,
Bar-Ilan University,
52900 Ramat-Gan, Israel
} \\
and \\
University of Luxembourg
\thanks{
Computer Science and Communications,
Faculty of Sciences,
6, rue Coudenhove-Kalergi,
L-1359 Luxembourg
} \\ \\
Karl Schlechta
\thanks{
ks@cmi.univ-mrs.fr, karl.schlechta@web.de, http://www.cmi.univ-mrs.fr/ $\sim$ ks
} \\
Laboratoire d'Informatique Fondamentale de Marseille
\thanks{
CMI, 39, rue Joliot-Curie, F-13453 Marseille Cedex 13, France
(UMR 7279, CNRS and Universit\'{e} de Provence, now Aix Marseille Universit\'{e})
}
}

\begin{document}

\newtheorem{lemma}{Lemma}[section]
\newtheorem{theorem}[lemma]{Theorem}
\newtheorem{proposition}[lemma]{Proposition}
\newtheorem{corollary}[lemma]{Corollary}
\newtheorem{claim}[lemma]{Claim}
\newtheorem{fact}[lemma]{Fact}
\newtheorem{remark}[lemma]{Remark}
\newtheorem{definition}{Definition}[section]
\newtheorem{construction}{Construction}[section]
\newtheorem{condition}{Condition}[section]
\newtheorem{example}{Example}[section]
\newtheorem{notation}{Notation}[section]
\newtheorem{bild}{Figure}[section]
\newtheorem{comment}{Comment}[section]
\newtheorem{statement}{Statement}[section]
\newtheorem{diagram}{Diagram}[section]

\renewcommand{\labelenumi}
  {(\arabic{enumi})}
\renewcommand{\labelenumii}
  {(\arabic{enumi}.\arabic{enumii})}
\renewcommand{\labelenumiii}
  {(\arabic{enumi}.\arabic{enumii}.\arabic{enumiii})}
\renewcommand{\labelenumiv}
  {(\arabic{enumi}.\arabic{enumii}.\arabic{enumiii}.\arabic{enumiv})}

\maketitle
\setcounter{secnumdepth}{4}
\setcounter{tocdepth}{4}

\tableofcontents

%
%
%
\chapter{
Introduction
}

We present here various results, which may one day be published in
a bigger paper, and which we wish to make already available to the
community.

We investigate several technical and conceptual questions.

Our main subject is the investigation of independence as a ternary
relation in the context of non-monotonic logic. In the context of
probability,
this investigation was started by W. Spohn et al., and then followed by
J. Pearl. We look at products of function sets, and thus continue our own
investigation of independence in non-monotonic logic. We show that a
finite characterization of this relation in our context is impossible, and
indicate how to construct all valid rules.
\chapter{
Countably many disjoint sets
}

We show here that - independent of the cardinality of the language -
one can define only countably many inconsistent formulas.

The question is due to D. Makinson (personal communication).

$ \xCO $

We show here that, independent of the cardinality of the language,
one can define only countably many inconsistent formulas.

The problem is due to D. Makinson (personal communication).

\be

$\hspace{0.01em}$


\label{Example Co-Ex-Inf}

There is a countably infinite set of formulas s.t. the defined model sets
are pairwise disjoint.

Let $p_{i}:i \xbe \xbo $ be propositional variables.

Consider $ \xbf_{i}:= \xcU \{ \xCN p_{j}:j<i\} \xcu p_{i}$ for $i \xbe
\xbo.$

Obviously, $M(\xbf_{i}) \xEd \xCQ $ for all $i.$

Let $i<i';$ we show $M(\xbf_{i}) \xcs M(\xbf_{i' })= \xCQ.$ $M(
\xbf_{i' }) \xcm \xCN p_{i},$ $M(\xbf_{i}) \xcm p_{i}.$

$ \xcz $
\\[3ex]

\ee

\bfa

$\hspace{0.01em}$


\label{Fact Co-Ex-Inf}

Any set $X$ of consistent formulas with pairwise disjoint model sets is at
most
countable \index{countable}

\efa

\subparagraph{
Proof
}

$\hspace{0.01em}$


Let such $X$ be given.

(1) We may assume that $X$ consists of conjunctions of propositional
variables
or their negations.

Proof: Rewrite all $ \xbf \xbe X$ as disjunctions of conjunctions $
\xbf_{j}.$ At least one of
the conjunctions $ \xbf_{j}$ is consistent. Replace $ \xbf $ by one such $
\xbf_{j}.$ Consistency
is preserved, as is pairwise disjointness.

(2) Let $X$ be such a set of formulas. Let $X_{i} \xcc X$ be the set of
formulas in $X$ with
length $i,$ i.e., a consistent conjunction of $i$ many propositional
variables or
their negations, $i>0.$

As the model sets for $X$ are pairwise disjoint, the model sets for all $
\xbf \xbe X_{i}$
have to be disjoint.

(3) It suffices now to show that each $X_{i}$ is at most countable; we
even show
that each $X_{i}$ is finite.

Proof by induction:

Consider $i=1.$ Let $ \xbf, \xbf' \xbe X_{1}.$ Let $ \xbf $ be $p$ or $
\xCN p.$ If $ \xbf' $ is not $ \xCN \xbf,$ then
$ \xbf $ and $ \xbf' $ have a common model. So one must be $p,$ the other
$ \xCN p.$ But these
are all possibilities, so $card(X_{1})$ is finite.

Let the result be shown for $k<i.$

Consider now $X_{i}.$ Take arbitrary $ \xbf \xbe X_{i}.$ Without loss of
generality, let
$ \xbf =p_{1} \xcu  \Xl  \xcu p_{i}.$ Take arbitrary
$ \xbf' \xEd \xbf.$ As $M(\xbf) \xcs M(\xbf')= \xCQ,$ $ \xbf' $
must be a conjunction containing one of
$ \xCN p_{k},$ $1 \xck k \xck i.$ Consider now $X_{i,k}:=\{ \xbf' \xbe
X_{i}: \xbf' $ contains $ \xCN p_{k}\}.$
Thus $X_{i}=\{ \xbf \} \xcv \xcV \{X_{i,k}:1 \xck k \xck i\}.$ Note that
all $ \xbq, \xbq' \xbe X_{i,k}$ agree on $ \xCN p_{k},$
so the situation in $X_{i,k}$ is isomorphic to $X_{i-1}.$ So,
by induction hypothesis, $card(X_{i,k})$ is finite,
as all $ \xbf' \xbe X_{i,k}$ have to be mutually inconsistent. Thus,
$card(X_{i})$ is finite.
(Note that we did not use the fact that elements from different $X_{i,k},$
$X_{i,k' }$
also have to be mutually inconsistent; our rough proof suffices.)

$ \xcz $
\\[3ex]

Note that the proof depends very little on logic. We needed normal forms,
and used two truth values. Obviously, we can easily generalize to finitely
many truth values.

$ \xCO $
\chapter{
Independence as ternary relation
}

$ \xCO $
\section{
Introduction
}

\label{Section Prob-Func}
\subsection{
Independence
}

Independence is a central concept of reasoning.

In the context of non-monotonic logic and related areas like theory
revision,
it was perhaps first investigated formally by R. Parikh and co-authors, see
e.g.  \cite{Par96}, to
obtain ``local'' conflict solution.

The present authors investigated its role for interpolation in
preferential logics in
 \cite{GS10}, and showed connections to abstract multiplication of
size.

Independence plays also a central role for a FOL treatment of
preferential logics, where problems like the
``dark haired Swedes'' have to be treated. This is still subject
of ongoing research.

J. Pearl investigated independence in graphs and pobabilistic reasoning,
e.g. in  \cite{Pea88}, also as a ternary relation, $ \xBc X \xfA Y
\xfA Z \xBe.$

The aim of the present paper is to extend this abstract approach to the
preferential situation. We should emphasize that this is only an abstract
description of the independence relation, and thus not the same as
independence
for non-monotonic interpolation as examined in
 \cite{GS10}, where we $ \xCf used$ independence, essentially in the
form of the
multiplicative law
$ \xbm (X \xCK Y)= \xbm (X) \xCK \xbm (Y),$ which says that the $ \xbm
-$function preserves independence.

We have not investigated if an interesting form of interpolation results
from
some application of $ \xbm $ to situations described by $ \xBc X \xfA Y
\xfA Z \xBe,$ analogously
to above application of $ \xbm $ to situations described by $ \xBc X \xfA
\xfA Y \xBe.$
\subsection{
Overview
}

We will first discuss simple examples, to introduce the main ideas.

We then present the basic definitions formally, for probabilistic and
set independence.

We then show basic results for set independence as a ternary relation, and
turn
to our main
results, absence of finite characterization, and construction of new rules
for
this ternary relation.
\subsection{
Discussion of some simple examples
}

\label{Section Discussion-Simple}

We consider here $X=Y=Z=W=\{0,1\}$ and their products.
We will later generalize, but the main ideas stay the same.
First, we look at $X \xCK Z$ (the Cartesian product of $X$ with $Z),$ then
at
$X \xCK Z \xCK W,$ at $X \xCK Y \xCK Z,$ finally at $X \xCK Y \xCK Z \xCK
W.$ Elements of these products,
i.e., sequences, will be written for simplicity 00, 01, 10, etc., context
will disambiguate. General sequences will often be written $ \xbs,$ $
\xbt,$ etc.
We will also look at subsets of these products, like $\{00,11\} \xcc X
\xCK Z,$ and
various probability measures on these products.

As a matter of fact, the main part of this article concerns subsets A
of products $X_{1} \xCK  \Xl  \xCK X_{n}$ and a suitable notion of
independence for A,
roughly, if we can write A as $A_{1} \xCK  \Xl  \xCK A_{m}.$ This will be
made more
precise and discussed in progressively more complicated cases in this
section.

In the context of preferential structures, A is intended to be
$ \xbm (X_{1} \xCK  \Xl  \xCK X_{n}),$ the set of minimal models of $X_{1}
\xCK  \Xl  \xCK X_{n}.$
\subsubsection{
$ X \xCK Z $
}

Let $P:X \xCK Z \xcp [0,1]$ be a (fixed) probability measure.

If $A \xcc X \xCK Z,$ we will set $P(A):= \xbS \{P(\xbs): \xbs \xbe
A\}.$

If $A_{x}:=\{ \xbs \xbe X \xCK Z: \xbs (X)=x\},$ we will write $P(x)$ for
$P(A_{x}),$ likewise
$P(z)$ for $P(A_{z}),$ if $A_{z}:=\{ \xbs \xbe X \xCK Z: \xbs (Z)=z\}.$
When these are ambiguous, we will e.g. write $A_{X=0}$ for $\{ \xbs \xbe X
\xCK Z: \xbs (X)=0\},$
and $P(X=0)$ for $P(A_{X=0}),$ etc.

We say that $X$ and $Z$ are independent for this $P$ iff for all
$xz \xbe X \xCK Z$ $P(xz)=P(x)*P(z).$

We write then $ \xBc X \xfA \xfA Z \xBe_{P},$
and call this and its variants probabilistic independence.

\be

$\hspace{0.01em}$


\label{Example XZ}

(1)

$P(00)=P(01)=1/6,$ $P(10)=P(11)=1/3.$

Then $P(X=0)=1/6+1/6=1/3,$ and $P(X=1)=2/3,$ $P(Z=0)=1/6+1/3=1/2,$ and
$P(Z=1)=1/2,$
so $ \xBc X \xfA \xfA Z \xBe_{P}.$

(2)

$P(00)=P(11)=1/3,$ $P(01)=P(10)=1/6.$

Then $P(X=0)=P(X=1)=P(Z=0)=P(Z=1)=1/2,$ but $P(00)=1/3 \xEd 1/2*1/2=1/4,$
so $ \xCN \xBc X \xfA \xfA Z \xBe_{P}.$

\ee

\bd

$\hspace{0.01em}$


\label{Definition P-A}

Consider now $ \xCQ \xEd A \xcc X \xCK Z$ for general $X,Z.$

Define the following probability measure on $X \xCK Z:$

\begin{flushleft}
\[  P_A(\xbs):=
\left\{ \begin{array}{lcl}
{\frac{1}{card(A)}}
\xEH iff \xEH  \xbs \xbe A  \xEP
\xEH \xEH \xEP
0 \xEH iff \xEH \xbs \xce A \xEP
\end{array}
\right.
\]
\end{flushleft}

\ed

\be

$\hspace{0.01em}$


\label{Example XZ-A}

(1)

$A:=\{00,01\},$

then $P_{A}(00)=P_{A}(01)=1/2,$ $P_{A}(10)=P_{A}(11)=0,$ $P_{A}(X=0)=1,$
$P_{A}(X=1)=0,$
$P_{A}(Z=0)=P_{A}(Z=1)=1/2,$ and we have $ \xBc X \xfA \xfA Z
\xBe_{P_{A}}.$

(2)

$A:=\{00,11\},$

then $P_{A}(00)=P_{A}(11)=1/2,$ $P_{A}(01)=P_{A}(10)=0,$
$P_{A}(X=0)=P_{A}(X=1)=1/2,$
$P_{A}(Z=0)=P_{A}(Z=1)=1/2,$ but $P_{A}(00)=1/2 \xEd
P_{A}(X=0)*P_{A}(Z=0)=1/4,$
and we have $ \xCN \xBc X \xfA \xfA Z \xBe_{P_{A}}.$

(3)

$A:=\{00,01,11\},$

then $P_{A}(00)=P_{A}(01)=P_{A}(11)=1/3,$ $P_{A}(10)=0,$ $P_{A}(X=0)=2/3,$
$P_{A}(X=1)=1/3,$
$P_{A}(Z=0)=1/3,$ $P_{A}(Z=1)=2/3,$
but $P_{A}(00)=1/3 \xEd P_{A}(X=0)*P_{A}(Z=0)=2/3*1/3=2/9,$
and we have $ \xCN \xBc X \xfA \xfA Z \xBe_{P_{A}}.$

\ee

Note that in (1) above, $A=\{0\} \xCK \{0,1\},$ but neither in (2), nor in
(3), A
can be written as such a product.
This is no coincidence, as we will see now.

More formally, we write $ \xBc X \xfA \xfA Z \xBe_{A}$ iff
for all $ \xbs \xbt \xbe A$ there is $ \xbr \xbe A$ such that $ \xbr (X)=
\xbs (X)$ and $ \xbr (Z)= \xbt (Z),$
or, equivalently, that $A=\{ \xbs (X): \xbs \xbe A\} \xCK \{ \xbs (Z):
\xbs \xbe A\},$
meaning that we can combine fragments of functions in A arbitrarily.

We call this and its variants set independence.

\bfa

$\hspace{0.01em}$


\label{Fact XZ-A}

Consider above situation $X \xCK Z.$ Then $ \xBc X \xfA \xfA Z
\xBe_{P_{A}}$ iff $ \xBc X \xfA \xfA Z \xBe_{A}.$

\efa

\subparagraph{
Proof
}

$\hspace{0.01em}$


``$ \xch $'':

$A \xcc \{ \xbs (X): \xbs \xbe A\} \xCK \{ \xbs (Z): \xbs \xbe A\}$ is
trivial.
Suppose $P_{A}(x,z)=P_{A}(x)*P_{A}(z),$ but there are $ \xbs, \xbt \xbe
A,$ $ \xbs (X) \xbt (Z) \xce A.$
Then $P_{A}(x),P_{A}(z)>0,$ but $P_{A}(x,z)=0,$ a contradiction.

``$ \xci $'':

Case 1: $P_{A}(x)=0,$ then $P_{A}(x,z)=0,$ and we are done. Likewise for
$P_{A}(Z)=0.$

Case 2: $P_{A}(x),P_{A}(z)>0.$

By definition and prerequisite,

$P_{A}(x)$ $=$ $ \frac{card\{ \xbs \xbe A: \xbs (X)=x\}}{card(A)}$ $=$ $
\frac{card\{ \xbs (Z): \xbs \xbe A\}}{card(A)},$

$P_{A}(z)$ $=$ $ \frac{card\{ \xbs \xbe A: \xbs (Z)=z\}}{card(A)}$ $=$ $
\frac{card\{ \xbs (X): \xbs \xbe A\}}{card(A)},$

$P_{A}(x,z)$ $=$ $ \frac{card\{ \xbs \xbe A: \xbs (X)=x, \xbs
(Z)=z\}}{card(A)}$ $=$ $ \frac{1}{card(A)}.$

By prerequisite again, $card(A)$ $=$ $card\{ \xbs (X): \xbs \xbe A\}$ $=$
$card\{ \xbs (Z): \xbs \xbe A\},$ so
$ \frac{card\{ \xbs (Z): \xbs \xbe A\}}{card(A)}$ $*$ $ \frac{card\{ \xbs
(X): \xbs \xbe A\}}{card(A)}$ $=$ $ \frac{1}{card(A)}$

$ \xcz $
\\[3ex]
\subsubsection{
$ X \xCK Z \xCK W $
}

Here, $W$ will not be mentioned directly.

Let $P:X \xCK Z \xCK W \xcp [0,1]$ be a probability measure.

Again, we say that $X$ and $Z$ are independent for $P,$ $ \xBc X \xfA \xfA
Z \xBe_{P},$ iff
for all $x \xbe X,$ $z \xbe Z$ $P(x,z)=P(x)*P(z).$

\be

$\hspace{0.01em}$


\label{Example XZW}

(1)

Let $P(000)=P(001)=P(010)=P(011)=1/12,$ $P(100)=P(101)=P(110)=P(111)=1/6,$
then
$X$ and $Z$ are independent.

(2)

Let $P(100)=P(101)=P(010)=P(011)=1/12,$ $P(000)=P(001)=P(110)=P(111)=1/6,$
then
$P(X=0)=P(X=1)=P(Z=0)=P(Z=1)=1/2,$ but $P(X=0,Z=0)=1/3 \xEd 1/2*1/2=1/4,$
so $ \xCN \xBc X \xfA \xfA Z \xBe_{P}.$

\ee

As above, we define $P_{A}$ for $ \xCQ \xEd A \xcc X \xCK Z \xCK W.$

\be

$\hspace{0.01em}$


\label{Example XZW-A}

(1)

$A:=\{000,001,010,011\}.$ Then $P_{A}(X=0,Z=0)=P_{A}(X=0,Z=1)=1/2,$
$P_{A}(X=1,Z=0)=P_{A}(X=1,Z=1)=0,$ $P_{A}(X=0)=1,$ $P_{A}(X=1)=0,$
$P_{A}(Z=0)=P_{A}(Z=1)=1/2,$ so
$X$ and $Z$ are independent.

(2)

For $A:=\{000,001,110,111\},$ we see that $X$ and $Z$ are not independent
for $P_{A}.$

\ee

Considering possible decompositions of A into set products, we are not so
much
interested how many continuations into $W$ we have, but if there are any
or none.
This is often the case in logic, we are not interested how many models
there
are, but if there is a model at all.

Thus we define independence for A again by:

$ \xBc X \xfA \xfA Z \xBe_{A}$ iff
for all $ \xbs \xbt \xbe A$ there is $ \xbr \xbe A$ such that $ \xbr (X)=
\xbs (X)$ and $ \xbr (Z)= \xbt (Z).$

The equivalence between probabilitistic independence,
$ \xBc X \xfA \xfA Z \xBe_{P_{A}}$ and set independence, $ \xBc X \xfA
\xfA Z \xBe_{A}$ is lost now, as the
second part of the
following example shows:

\be

$\hspace{0.01em}$


\label{Example XZW-A-2}

(1)

$A:=\{000,010,100,110\}$ satisfies both forms of independence,
$ \xBc X \xfA \xfA Z \xBe_{P_{A}}$ and set independence, $ \xBc X \xfA
\xfA Z \xBe_{A}.$

\ee

(2)

$A:=\{000,001,010,100,110\}.$

Here, we have $P_{A}(X=0)=3/5,$ $P_{A}(X=1)=2/5,$ $P_{A}(Z=0)=3/5,$
$P_{A}(Z=1)=2/5,$
but $P_{A}(X=0,Z=0)=2/5 \xEd 3/5*3/5.$

Consider now $ \xBc X \xfA \xfA Z \xBe_{A}:$ Take $ \xbs, \xbt \xbe A,$
then for all possible values
$ \xbs (X),$ $ \xbt (Z),$ there is $ \xbr $ such that $ \xbr (X)= \xbs
(X),$ $ \xbr (Z)= \xbt (Z)$ - the value
$ \xbr (W)$ is without importance.

We have, however:

\bfa

$\hspace{0.01em}$


\label{Fact XZW-A}

$ \xBc X \xfA \xfA Z \xBe_{P_{A}}$ $ \xch $ $ \xBc X \xfA \xfA Z
\xBe_{A}.$

\efa

\subparagraph{
Proof
}

$\hspace{0.01em}$


Let $ \xbs, \xbt \xbe A,$ but suppose there is no $ \xbr \xbe A$ such
that $ \xbr (X)= \xbs (X)$ and
$ \xbr (Z)= \xbt (Z).$ Then $P_{A}(\xbs (X)),P_{A}(\xbt (Z))>0,$ but
$P_{A}(\xbs (X), \xbt (Z))=0.$ $ \xcz $
\\[3ex]
\subsubsection{
$ X \xCK Y \xCK Z $
}

\label{Section XYZ}

We consider now independence of $X$ and $Z,$ given $Y.$

The probabilistic definition is:

$ \xBc X \xfA Y \xfA Z \xBe_{P}$ iff for all $x \xbe X,y \xbe Y,z \xbe Z$
$P(x,y,z)*P(y)=P(x,y)*P(y,z).$

As we are interested mainly in subsets $A \xcc X \xCK Y \xCK Z$ and the
resulting $P_{A},$
and combination of function fragments, we work immediately with these.

We have to define $ \xBc X \xfA Y \xfA Z \xBe_{A}.$

$ \xBc X \xfA Y \xfA Z \xBe_{A}$ iff for all $ \xbs, \xbt \xbe A$ such
that $ \xbs (Y)= \xbt (Y)$ there is $ \xbr \xbe A$
such that $ \xbr (X)= \xbs (X),$ $ \xbr (Y)= \xbs (Y)= \xbt (Y),$ $ \xbr
(Z)= \xbt (Z).$

When we set for $y \xbe Y$ $A_{y}:=\{ \xbs \xbe A: \xbs (Y)=y\},$ we then
have:

$A_{y}=\{ \xbs (X): \xbs \xbe A_{y}\} \xCK \{y\} \xCK \{ \xbs (Z): \xbs
\xbe A_{y}\}.$

The following example shows that $ \xBc X \xfA Y \xfA Z \xBe_{A}$ and $
\xBc X \xfA \xfA Z \xBe_{A}$ are independent
from each other:

\be

$\hspace{0.01em}$


\label{Example XYZ-A}

(1)

$ \xBc X \xfA Y \xfA Z \xBe_{A}$ may hold, but not $ \xBc X \xfA \xfA Z
\xBe_{A}:$

Consider $A:=\{000,111\}.$ $ \xBc X \xfA Y \xfA Z \xBe_{A}$ is obvious, as
only $ \xbs $ goes through
each element in the middle. But there is no 0x1, so $ \xBc X \xfA \xfA Z
\xBe_{A}$ fails.

(2)

$ \xBc X \xfA \xfA Z \xBe_{A}$ may hold, but not $ \xBc X \xfA Y \xfA Z
\xBe_{A}:$

Consider $A:=\{000,101,110,011\}.$ Fixing, e.g., 0 in the middle shows
that
$ \xBc X \xfA Y \xfA Z \xBe_{A}$ fails, but neglecting the middle, we can
combine arbitrarily, so
$ \xBc X \xfA \xfA Z \xBe_{A}$ holds.

\ee

\be

$\hspace{0.01em}$


\label{Example XYZ-Prod}

This example show that $ \xBc X \xfA Y \xfA Z \xBe_{A}$ does not mean that
A is some
product $A_{X} \xCK A_{Y} \xCK A_{Z}:$

Let $A:=\{000,111\},$ then clearly $ \xBc X \xfA Y \xfA Z \xBe_{A},$ but A
is no such product.

\ee

We have again:

\bfa

$\hspace{0.01em}$


\label{Fact XYZ-A}

Let $ \xCQ \xEd A \xcc X \xCK Y \xCK Z,$ then
$ \xBc X \xfA Y \xfA Z \xBe_{A}$ and $ \xBc X \xfA Y \xfA Z \xBe_{P_{A}}$
are equivalent.

\efa

\subparagraph{
Proof
}

$\hspace{0.01em}$


``$ \xci $'':

Suppose there are $ \xbs, \xbt \xbe A$ such that $ \xbs (Y)= \xbt (Y),$
but there is no $ \xbr \xbe A$
such that $ \xbr (X)= \xbs (X),$ $ \xbr (Y)= \xbs (Y)= \xbt (Y),$ $ \xbr
(Z)= \xbt (Z).$
Then $P_{A}(\xbs (X), \xbs (Y)),P_{A}(\xbt (Y), \xbt (Z)),P_{A}(\xbs
(Y))>0,$ but
$P_{A}(\xbs (X), \xbs (Y)= \xbt (Y), \xbt (Z))=0.$

``$ \xch $'':

Case 1: $P_{A}(x,y)$ or $P_{A}(y,z)=0,$ then $P_{A}(x,y,z)=0,$ and we are
done.

Case 2: $P_{A}(x,y),P_{A}(y,z)>0.$
By definition and prerequisite,
$P_{A}(x,y)$ $=$ $ \frac{card\{ \xbs \xbe A: \xbs (X)=x, \xbs
(Y)=y\}}{card(A)}$ $=$ $ \frac{card\{ \xbs (Z): \xbs \xbe A, \xbs
(Y)=y\}}{card(A)}$ and
$P_{A}(y,z)$ $=$ $ \frac{card\{ \xbs \xbe A: \xbs (Y)=y, \xbs
(Z)=Z\}}{card(A)}$ $=$ $ \frac{card\{ \xbs (X): \xbs \xbe A, \xbs
(Y)=y\}}{card(A)},$ so
$P_{A}(x,y)*P_{A}(y,z)$ $=$ $ \frac{card\{ \xbs \xbe A: \xbs
(Y)=y\}}{card(A)*card(A)}.$ Moreover,
$P_{A}(y)$ $=$ $ \frac{card\{ \xbs \xbe A: \xbs (Y)=y\}}{card(A)},$
$P_{A}(x,y,z)$ $=$ $ \frac{1}{card(A)},$ so
$P_{A}(y)*P_{A}(x,y,z)$ $=$ $ \frac{card\{ \xbs \xbe A: \xbs
(Y)=y\}}{card(A)*card(A)}$ $=$ $P_{A}(x,y)*P_{A}(y,z)$

$ \xcz $
\\[3ex]
\subsubsection{
$ X \xCK Y \xCK Z \xCK W $
}

The definitions stay the same as for $X \xCK Y \xCK Z.$

The equivalence between probabilitistic independence,
$ \xBc X \xfA Y \xfA Z \xBe_{P_{A}}$ and set independence, $ \xBc X \xfA Y
\xfA Z \xBe_{A}$ is lost again, as the
following example shows:

\be

$\hspace{0.01em}$


\label{Example XYZW-A}

$A:=\{0000,0001,0010,1000,1010\}.$

Here, we have $P_{A}(X=0,Y=0)=3/5,$ $P_{A}(X=1,Y=0)=2/5,$
$P_{A}(Y=0,Z=0)=3/5,$ $P_{A}(Y=0,Z=1)=2/5,$ $P_{A}(Y=0)=1,$
but $P_{A}(X=0,Y=0,Z=0)=2/5 \xEd 3/5*3/5.$

Consider now $ \xBc X \xfA Y \xfA Z \xBe_{A}:$ Take $ \xbs, \xbt \xbe A,$
such that $ \xbs (Y)= \xbt (Y),$
then for all possible values
$ \xbs (X),$ $ \xbt (Z),$ there is $ \xbr $ such that $ \xbr (X)= \xbs
(X),$ $ \xbr (Y)= \xbs (Y)= \xbt (Y),$
$ \xbr (Z)= \xbt (Z)$ - the value $ \xbr (W)$ is without importance.

\ee

We have, however:

\bfa

$\hspace{0.01em}$


\label{Fact XYZW-A}

$ \xBc X \xfA Y \xfA Z \xBe_{P_{A}}$ $ \xch $ $ \xBc X \xfA Y \xfA Z
\xBe_{A}.$

\efa

\subparagraph{
Proof
}

$\hspace{0.01em}$


Let $ \xbs, \xbt \xbe A$ such that $ \xbs (Y)= \xbt (Y),$ but suppose
there is no $ \xbr \xbe A$ such that
$ \xbr (X)= \xbs (X),$ $ \xbr (Y)= \xbs (Y)= \xbt (Y),$
$ \xbr (Z)= \xbt (Z).$ Then $P_{A}(\xbs (X), \xbs (Y)),P_{A}(\xbs (Y),
\xbt (Z))>0,$
but $P_{A}(\xbs (X), \xbs (Y), \xbt (Z))=0.$ $ \xcz $
\\[3ex]
\subsubsection{
A remark on generalization
}

The $X,Y,Z,W$ may also be more complicated sets, themselves products,
but this will not change definitions and results beyond notation.

In the more complicated cases, we will often denote subsets by more
complicated letters than A, e.g., by $ \xbS.$
\subsubsection{
A remark on intuition
}

Consider set independence, where $A:= \xbm (U),$ $U=U_{1} \xCK  \Xl  \xCK
U_{n}.$
Set $ \xBc  \Xl  \xBe:= \xBc  \Xl  \xBe_{ \xbm (U)}.$

 \xEh

 \xDH
$ \xBc X \xfA \xfA Z \xBe $ means then:

 \xEh
 \xDH
all we know is that we are in a normal situation,
 \xDH
if we know in addition something definite about $Z$ (1 model!)
we do not know anything more about $X,$ and vice versa.
 \xEj

$ \xBc X \xfA Y \xfA Z \xBe $ means then:

 \xEh
 \xDH
all we know is that we are in a normal situation,
 \xDH
if we have definite information about $Y,$
we may know more about $X.$ But knowing something
in addition about $Z$ will not give us not more information about $X,$ and
conversely.
 \xEj

 \xDH
The restriction to $ \xbm (U)$ codes our background knowledge.

 \xDH
Note that $X \xcv Y \xcv Z$ need not be $I,$ e.g., $W$ might be missing.
We did not
count the continuations into $W,$ but considered only existence of a
continuation
(if this does not exist, then there just is no such sequence).

This corrsponds to multiplication with 1, the unit ALL on $W,$ or, more
generally, in the rest of the paper, with $1_{I-(X \xcv Y \xcv Z)}.$
We may choose however
we want, it has to be somewhere, in ALL.

 \xEj
\subsection{
Basic definitions
}

\bd

$\hspace{0.01em}$


\label{Definition Restrict}

If $f$ is a function, $Y$ a subset of its domain, we write
$f \xex Y$ for the restriction of $f$ to elements of $Y.$

If $F$ is a set of functions over $Y,$ then $F \xex Y:=\{f \xex Y:f \xbe
F\}.$
\section{
Probabilistic and set independence
}
\subsection{
Probabilistic independence
}

\ed

Independence as an abstract ternary relation for probability and other
situations has been examined by W. Spohn,
see  \cite{Spo80}, A. P. Dawid,
see  \cite{Daw79},
J. Pearl, see, e.g.,
 \cite{Pea88}, etc.

\bd

$\hspace{0.01em}$


\label{Definition Restrict-2}

(1)

Let $I \xEd \xCQ $ be an arbitrary (index) set, for $i \xbe I$ $U_{i} \xEd
\xCQ $ arbitrary sets.
Let $U:= \xbP \{U_{i}:i \xbe I\},$ and for $X \xcc I$ $U_{X}:= \xbP
\{U_{i}:i \xbe X\}.$

(2)

Let $P: \xdp (U) \xcp [0,1]$ be a probability measure. (We may assume that
$P$ is defined
by its value on singletons.)

(3.1)

By abuse of language, for $X \xcc I,$ $x \xbe U_{X},$
let $P(x)$ $:=$ $P(\{u \xbe U: \xcA i \xbe Xu(i)=x(i)\}),$ so
$P(x)=P(\{u \xbe U:u \xex X=x\}).$

Analogously, for $X,Y \xcc I,$ $X \xcs Y= \xCQ,$ $x \xbe U_{X},$ $y \xbe
U_{Y},$
let $P(x,y)$ $:=$ $P(\{u \xbe U:$ $u \xex X=x$ and $u \xex Y=y\}).$

(3.2)

Finally, for $X,Y,Z \xcc I$ pairwise disjoint, $x \xbe U_{X},$ $y \xbe
U_{Y},$ $z \xbe U_{Z},$ let
$P(x \xfA y):= \frac{P(x,y)}{P(y)},$ $P(x \xfA y,z):=
\frac{P(x,y,z)}{P(y,z)},$ etc.

(We have, of course, to pay attention that we do not divide by 0.)

\ed

\bd

$\hspace{0.01em}$


\label{Definition Prob-Ind}

$P$ as above defines a 3-place relation of independence
on pairwise disjoint $X,Y,Z \xcc I$
$ \xBc X \xfA Y \xfA Z \xBe_{P}$ by

\begin{flushleft}
\[  \xBc  \xdx \xfA \xdy \xfA \xdz  \xBe _{P}:\xcr
\left\{ \begin{array}{lcl}
 \xcA x \xbe U_X,  \xcA y \xbe U_Y,  \xcA z \xbe U_Z
(P(y,z)> 0  \xcp  P(x \xfA y) =P(x \xfA y,z)),
\xEH if \xEH  Y \xEd \xCQ  \xEP
i.e., P(x,y)/P(y)=P(x,y,z)/P(y,z), or
\xEH \xEH \xEP
P(x,y,z)*P(y)=P(x,y)*P(y,z)
\xEH \xEH \xEP
\xEH \xEH \xEP
\xEH \xEH \xEP
 \xcA x \xbe U_X,  \xcA z \xbe U_Z  (P(z)>0  \xcp
P(x)= P(x \xfA z)),
\xEH if \xEH  Y  = \xCQ  \xEP
i.e., P(x)=P(x,z)/P(z), or
\xEH \xEH \xEP
P(x,z)=P(x)*P(z)
\xEH \xEH \xEP
\end{array}
\right.
\]
\end{flushleft}

If $Y= \xCQ,$ we shall also write $ \xBc X \xfA \xfA Z \xBe_{P}$ for $
\xBc X \xfA Y \xfA Z \xBe_{P}$.

Recall from
Section 
\ref{Section Discussion-Simple} (page 
\pageref{Section Discussion-Simple})  that we call
this notion probabilistic independence.

\ed

E.g., Pearl discusses the rules $(a)-(e)$ of
Definition 
\ref{Definition Basic-Rules} (page 
\pageref{Definition Basic-Rules})  for the relation
defined in Definition 
\ref{Definition Prob-Ind} (page 
\pageref{Definition Prob-Ind}).

\bd

$\hspace{0.01em}$


\label{Definition Basic-Rules}

(a) Symmetry: $ \xBc X \xfA Y \xfA Z \xBe $ $ \xcr $ $ \xBc Z \xfA Y \xfA
X \xBe $

(b) Decomposition: $ \xBc X \xfA Y \xfA Z \xcv W \xBe $ $ \xcp $ $ \xBc X
\xfA Y \xfA Z \xBe $

(c) Weak Union: $ \xBc X \xfA Y \xfA Z \xcv W \xBe $ $ \xcp $ $ \xBc X
\xfA Y \xcv W \xfA Z \xBe $

(d) Contraction: $ \xBc X \xfA Y \xfA Z \xBe $ and $ \xBc X \xfA Y \xcv Z
\xfA W \xBe $ $ \xcp $ $ \xBc X \xfA Y \xfA Z \xcv W \xBe $

(e) Intersection: $ \xBc X \xfA Y \xcv W \xfA Z \xBe $ and $ \xBc X \xfA
Y \xcv Z \xfA W \xBe $ $ \xcp $ $ \xBc X \xfA Y \xfA Z \xcv W \xBe $

$(\xCQ)$ Empty outside: $ \xBc X \xfA Y \xfA Z \xBe $ if $X= \xCQ $ or
$Z= \xCQ.$

\ed

\bp

$\hspace{0.01em}$


\label{Proposition Prob-Valid}

If $P$ is a probability measure, and $ \xBc X \xfA Y \xfA Z \xBe_{P}$
defined as above, then
$(a)-(d)$ of Definition 
\ref{Definition Basic-Rules} (page 
\pageref{Definition Basic-Rules})  hold for $ \xBc  \Xl
\xBe = \xBc  \Xl  \xBe_{P},$
and if $P$ is strictly positive,
(e) will also hold.

\ep

The proof is elementary, well known, and will not be repeated here.

Doch ein Beispiel geben?
\subsubsection{
A side remark on preferential structures
}

Being a minimal element is not upward absolute in general preferential
structures, but in raked structures, provided the smaller set contains
some element minimal in the bigger set.

\bfa

$\hspace{0.01em}$


\label{Fact Basic-Pro}

In the probabilistic interpretation, the following holds:

Let $U$ be a finite set, $f:U \xcp \xdR $ such that $ \xcA u \xbe U.f(u)
\xcg 0.$

For all $A \xcc U,$ such that $ \xcE a' \xbe A.f(a')>0$ and all $a \xbe
A$

$f_{A}(a):= \frac{f(a)}{ \xbS \{f(a'):a' \xbe A\}}$ defines a probability
measure on $ \xCf A.$

For $B \xcc A,$ define $f_{A}(B):= \xbS \{f_{A}(b):b \xbe B\}.$ Then the
following property
holds:

(BASIC) For all $D \xcc B \xcc A \xcc U$ such that $ \xcE b \xbe B.f(b)>0$
$f_{A}(D)=f_{A}(B)*f_{B}(D).$

\efa

\subparagraph{
Proof
}

$\hspace{0.01em}$


For $X \xcc Y \xcc U$ such that $ \xcE y \xbe Y.f(y)>0$ we have
$f_{Y}(X):= \xbS \{f_{Y}(x):x \xbe X\}= \frac{ \xbS \{f(x):x \xbe X\}}{
\xbS \{f(y):y \xbe Y\}}.$

Thus, $f_{A}(D)$ $:=$ $ \frac{ \xbS \{f(d):d \xbe D\}}{ \xbS \{f(a):a \xbe
A\}}$ $=$ $ \frac{ \xbS \{f(b):b \xbe B\}}{ \xbS \{f(a):a \xbe A\}}$ $*$ $
\frac{ \xbS \{f(d):d \xbe D\}}{ \xbS \{f(b):b \xbe
B\}}$ $=$ $f_{A}(B)*f_{B}(D).$

$ \xcz $
\\[3ex]

We have the following fact for $ \xbm $ generated by a relation:

\bfa

$\hspace{0.01em}$


\label{Fact Basic-Pref}

Let $U$ be a finite preferential structure such that for $A \xcc U$ $ \xbm
(A)= \xCQ $ $ \xch $ $A= \xCQ.$

Then $U$ is ranked iff (BASIC) as defined in
Fact \ref{Fact Basic-Pro} (page \pageref{Fact Basic-Pro})  holds for $f_{A}.$

\efa

\subparagraph{
Proof
}

$\hspace{0.01em}$


``$ \xch $'':

Let $D \xcc B \xcc A \xcc U,$ $B \xEd \xCQ.$

Case 1: $D \xcs \xbm (A)= \xCQ.$ Then $f_{A}(D)=0.$

Case 1.1: If $B \xcs \xbm (A)= \xCQ,$ then $f_{A}(B)=0,$ and we are done.

Case 1.2: Let $B \xcs \xbm (A) \xEd \xCQ.$ If $D \xcs \xbm (B)= \xCQ,$
then $f_{B}(D)=0,$ and we are done.
Suppose $D \xcs \xbm (B) \xEd \xCQ,$ so there is $d \xbe D \xcs \xbm
(B),$
so $d \xbe D \xcs \xbm (A)$ by $B \xcs \xbm (A) \xEd \xCQ $ and
rankedness, so $f_{A}(D) \xEd \xCQ,$
contradiction.

Case 2: $D \xcs \xbm (A) \xEd \xCQ.$

Thus, by $D \xcc B,$ $B \xcs \xbm (A) \xEd \xCQ,$ and by rankedness $
\xbm (B)=B \xcs \xbm (A).$ So
by $D \xcc B$ again, $D \xcs \xbm (A)=D \xcs (B \xcs \xbm (A))=D \xcs \xbm
(B).$
By definition,
$f_{A}(B):= \frac{card(\xbm (A) \xcs B)}{card(\xbm (A))},$
$f_{A}(D):= \frac{card(\xbm (A) \xcs D)}{card(\xbm (A))},$
$f_{B}(D):= \frac{card(\xbm (B) \xcs D)}{card(\xbm (B))}.$
Thus,
$ \frac{card(\xbm (A) \xcs D)}{card(\xbm (A))}= \frac{card(\xbm (A)
\xcs B)}{card(\xbm (A))}* \frac{card(\xbm (B) \xcs D)}{card(\xbm
(B))}.$

``$ \xci $'':

Then there are $a,b,c \xbe U,$ where $ \xCf a$ is incomparable to $b,$ and
$b \xeb c$ but $a \xeB c,$
or $c \xeb b,$ but $c \xeB a.$ We have four possible cases.

Let, in all cases, $A:=\{a,b,c\}.$ We construct a contradiction to
(BASIC).

Case 1, $b \xeb c:$

Case 1.1, a is incomparable to $c:$
Consider $B:=\{a,c\},$ $D:=\{a\}.$
Then $f_{A}(D)= \frac{1}{2},$ $f_{A}(B)= \frac{1}{2},$ $f_{B}(D)=
\frac{1}{2}.$

Case 1.2, $c \xeb a$ (so $ \xeb $ is not transitive):
Consider $B:=\{a,b\},$ $D:=\{a\}.$
Then $f_{A}(D)=0,$ $f_{A}(B)=1,$ $f_{B}(D)= \frac{1}{2}.$

Case 2, $c \xeb b:$

Case 2.1, a is incomparable to $c:$

Consider $B:=\{a,b\},$ $D:=\{a\}.$ Then
$f_{A}(D)= \frac{1}{2},$ $f_{A}(B)= \frac{1}{2},$ $f_{B}(D)= \frac{1}{2}.$

Case 2.2, $a \xeb c$ - similar to Case 1.2.

$ \xcz $
\\[3ex]

\br

$\hspace{0.01em}$


\label{Remark Zero}

Note that sets $A \xcc B,$ where $ \xbm (B) \xcs A= \xCQ,$ and sets where
$P(A)=0$ have a similar,
exceptional role. This might still be important.
\subsection{
Set independence
}

\er

We interpret independence here differently, but in a related way,
as prepared in
Section 
\ref{Section Discussion-Simple} (page 
\pageref{Section Discussion-Simple}).

\bd

$\hspace{0.01em}$


\label{Definition Indep-Fct}

We consider function sets $ \xbS $ etc. over a fixed, arbitrary domain $I
\xEd \xCQ,$ into
some fixed codomain $K.$

(1)

For pairwise disjoint subsets $X,Y,Z$ of $I,$ we define

$ \xBc X \xfA Y \xfA Z \xBe _{ \xbS }$ iff for all $f,g \xbe \xbS $ such that $f
\xex Y=g \xex Y,$
there is $h \xbe \xbS $ such that $h \xex X=f \xex X,$ $h \xex Y=f \xex
Y=g \xex Y,$
$h \xex Z=g \xex Z.$

Recall from
Section 
\ref{Section Discussion-Simple} (page 
\pageref{Section Discussion-Simple})  that we call
this notion set independence.

$Y$ may be empty, then the condition $f \xex Y=g \xex Y$ is void.

Note that nothing is said about $I-(X \xcv Y \xcv Z),$ so we look at the
projection
of $U$ to $X \xcv Y \xcv Z.$

When $Y= \xCQ,$ we will also write $ \xBc X \xfA \xfA Z \xBe _{ \xbS }.$

$ \xBc X \xfA Y \xfA Z \xBe _{ \xbS }$ means thus, that we can piece functions
together, or that we have a
sort of decomposition of $ \xbS $ into a product. This is an independence
property,
we can put parts together independently.

(2)

In the sequel, we will just write $ \xBc  \Xl  \xBe $ for $ \xBc  \Xl
\xBe_{ \xbS }$ when the meaning is
clear from the context.

\ed

Recall that
Example 
\ref{Example XZW-A-2} (page 
\pageref{Example XZW-A-2})  compares different forms of independence,
the
probabilistic and the set variant.

Obviously, we can generalize the equivalence results for probabilistic
and set independence for $X \xCK Z$ and $X \xCK Y \xCK Z$ to the general
situation with $W$ in
Section 
\ref{Section Discussion-Simple} (page 
\pageref{Section Discussion-Simple}),
as long as we do not consider the full functions $ \xbs,$ but only their
restrictions to $X,Y,Z,$ $ \xbs \xex (X \xcv Y \xcv Z).$
As we will stop the discussion of probablistic independece here, and
restrict ourselves to set independence, this is left as an easy exercise
to the reader.
\clearpage
\section{
Basic results for set independence
}

\bn

$\hspace{0.01em}$


\label{Notation Indep-Fct}

In more complicated cases,
we will often write $ \xCf ABC$ for $ \xBc A \xfA B \xfA C \xBe,$ and $ \xCN
ABC$
or $- \xCf ABC$ if $ \xBc A \xfA B \xfA C \xBe $ does
not hold.
Moreover, we will often just write $f(A)$ for $f \xex A,$ etc.

For $ \xBc A \xcv A' \xfA B \xfA C \xBe,$ we will then write $(AA')BC,$ etc.

If only singletons are involved, we will sometimes write $ \xCf abc$
instead of $ \xCf ABC,$
etc.

When we speak about fragments of functions, we will often write just
$A: \xbs $ for $ \xbs \xex A,$ $B: \xbs = \xbt $ for $ \xbs \xex B= \xbt
\xex B,$ etc.

\en

We use the following notations for functions:

\bd

$\hspace{0.01em}$


\label{Definition Const-Func-H}

The constant functions $0_{c}$ and $1_{c}:$

$0_{c}(i)=0$ for all $i \xbe I$

$1_{c}(i)=1$ for all $i \xbe I$

Moreover, when we define a function $ \xbs:I \xcp \{0,1\}$ argument by
argument,
we abbreviate $ \xbs (a)=0$ by $a=0,$ etc.

Sometimes, we also give (a fragment of) a function just by the sequence of
the
values, so instead of writing $a=0,$ $b=1,$ $c=1,$ we just write 011 -
context
will disambiguate.

\ed

\br

$\hspace{0.01em}$


\label{Remark Sys-Valid-H}

This remark gives an intuitive justification of (some of) above rules in
our context.

Rule (a) is trivial.

It is easiest to set $Y:= \xCQ $ to see the intuitive meaning.

Rule (b) is a trivial consequence. If we can combine longer sequences,
then we
can combine shorter, too.

Rule (c) is again a trivial consequence. If we can combine arbitrary
sequences,
then we can also combine those which agree already on some part.

Rule (d) is the most interesting one, it says when we may combine $ \xCf
longer$
sequences. Having just $ \xBc X \xfA \xfA Z \xBe $ and $ \xBc X \xfA \xfA
W \xBe $ as prerequisite does not
suffice, as we might lose when applying $ \xBc X \xfA \xfA W \xBe $ what
we had already by
$ \xBc X \xfA \xfA Z \xBe.$ The condition $ \xBc X \xfA Z \xfA W \xBe $
guarantees that we do not lose this.

In our context, it means the following:

We want to combine $ \xbs \xex X$ with $ \xbt \xex Z \xcv W.$
By $ \xBc X \xfA \xfA Z \xBe,$ we can combine $ \xbs \xex X$ with $ \xbt
\xex Z.$ Fix $ \xbr $ such that
$ \xbr \xex X= \xbs \xex X,$ $ \xbr \xex Z= \xbt \xex Z.$ As $ \xbr \xex
Z= \xbt \xex Z,$ by $ \xBc X \xfA Z \xfA W \xBe,$ we can
combine $ \xbr \xex X \xcv Z$ with $ \xbt \xex W,$ and have the result.

Note that we change the functions here, too: we start with $ \xbs,$ $
\xbt,$ then continue
with $ \xbr,$ $ \xbt.$

We can use what we constructed already as a sort of scaffolding for
constructing the rest.

\er

\bfa

$\hspace{0.01em}$


\label{Fact Sys-Prod}

Zusammenhang $ \xBc X \xfA Y \xfA Z \xBe $ mit Produkten.

\efa

\subparagraph{
Proof
}

$\hspace{0.01em}$


Do

$ \xcz $
\\[3ex]

We show now that above Rules $(a)-(d)$ hold in our context, but (e) does
not hold.

\bfa

$\hspace{0.01em}$


\label{Fact Sys-Valid-H}

In our interpretation,

(1)
rule (e) does not hold,

(2)
all $ \xBc X \xfA Y \xfA \xCQ \xBe $ (and thus also all $ \xBc \xCQ \xfA Y
\xfA Z \xBe)$ hold.

(3)
rules $(a)-(d)$ hold, even when one or both of the outside elements of the
tripels is the empty set.

\efa

\subparagraph{
Proof
}

$\hspace{0.01em}$


(1) (e) does not hold:

Consider $I:=\{x,y,z,w\}$ and
$U:=\{1111,0100\}.$ Then $x(yw)z$ and $x(yz)w,$
as for all $ \xbs \xex yw$ there is just one $ \xbt $
this $ \xbs $ can be. The same holds for $x(yz)w.$
But for $y=1,$ there are two different paths through $y=1,$ which cannot
be
combined.

(2) This is a trivial consequence of the fact that $\{f:$ $f: \xCQ \xcp
U\}=\{ \xCQ \}.$

(3)
Rules (a), (b), (c) are trivial, by definition, also for $X,Z= \xCQ.$
In (c), if $W= \xCQ,$ there is nothing to show.

Rule (d): The cases for $X,W,Z= \xCQ $ are trivial.
Assume $ \xbs,$ $ \xbt $ such that $ \xbs \xex Y= \xbt \xex Y,$ we want
to combine $ \xbs \xex X$ with
$ \xbt \xex Z \xcv W.$ By $ \xBc X \xfA Y \xfA Z \xBe,$ there is $ \xbr $
such that $ \xbr \xex X= \xbs \xex X,$ $ \xbr \xex Y= \xbs \xex Y= \xbt
\xex Y,$
$ \xba \xex X= \xbr \xex Z= \xbt \xex Z.$ Thus $ \xbr $ and $ \xbt $
satisfy the prerequisite of
$ \xBc X \xfA Y \xcv Z \xfA W \xBe,$ and there is $ \xba $ such that $
\xba \xex X= \xbr \xex X= \xbs \xex X,$
$ \xba \xex X= \xbr \xex Y= \xbs \xex Y= \xbt \xex Y,$ $ \xba \xex W= \xbt
\xex W.$

$ \xcz $
\\[3ex]

Next, we give examples which shows that increasing the center set
can change validity of the tripel in any way.

\be

$\hspace{0.01em}$


\label{Example Change-Rel}

(1)

This example shows that neither $ \xBc X \xfA Y \xfA Z \xBe $ implies $
\xBc X \xfA \xfA Z \xBe,$ nor, conversely,
$ \xBc X \xfA \xfA Z \xBe $ implies $ \xBc X \xfA Y \xfA Z \xBe.$

Consider $I:=\{x,y,z\}.$

(1.1) Let $U:=\{ \xBc 0,0,0 \xBe, \xBc 1,1,1 \xBe, \xBc 0,1,0 \xBe,
\xBc 1,0,1 \xBe, \xBc 1,1,0 \xBe, \xBc 0,0,1 \xBe \}.$
Then $ \xBc x \xfA \xfA z \xBe,$ as all combinations for $x$ and $y$
exist, i.e. paths with
the projections $ \xBc 0,0 \xBe,$ $ \xBc 0,1 \xBe,$ $ \xBc 1,0 \xBe,$ $
\xBc 1,1 \xBe.$
Fix, e.g., $y=1.$ Then the paths through $y=1$ are
$ \xBc 1,1,1 \xBe,$ $ \xBc 0,1,0 \xBe,$ $ \xBc 1,1,0 \xBe,$ but $ \xBc
0,1,1 \xBe $ is missing. So
$ \xBc x \xfA y \xfA z \xBe $ does not hold.

(1.2) Let $U:=\{ \xBc 0,0,0 \xBe, \xBc 1,1,1 \xBe \}.$ Then $ \xBc x \xfA
\xfA z \xBe $ trivially fails, but
$ \xBc x \xfA y \xfA z \xBe $ holds.

(2)

Consider $I:=\{x,a,b,c,d,z\}.$

Let $ \xbS:=\{111111,$ 011110, 011101, 111100, 110111, $010000\}.$

Then $ \xCN x(abcd)z,$ $x(abc)z,$ $ \xCN x(ab)z.$

For $ \xCN x(abcd)z,$ fix $abcd=1111,$ then $111111,011110 \xbe \xbS,$
but, e.g., $011111 \xce \xbS.$

For $x(abc)z,$ the following combinations of abc exist: $111,101,100.$
The result is trivial for 101 and 100. For 111, all combinations for $x$
and $z$
with 0 and 1 exist.

For $ \xCN x(ab)z,$ fix $ab=10,$ then $110111,010000 \xbe \xbS,$ but
there is, e.g., no
$110xy0 \xce \xbS.$

See Diagram \ref{Diagram Change-Rel} (page \pageref{Diagram Change-Rel})

$ \xcz $
\\[3ex]

\ee

$ \xCO $

\vspace{10mm}

\begin{diagram}

\label{Diagram Change-Rel}

\centering
\setlength{\unitlength}{1mm}
{\renewcommand{\dashlinestretch}{30}
\begin{picture}(160,190)(0,0)

\put(10,170){$x$}
\put(30,170){$a$}
\put(50,170){$b$}
\put(70,170){$c$}
\put(90,170){$d$}
\put(110,170){$z$}

\put(10,167){\circle*{1}}
\put(30,167){\circle*{1}}
\put(50,167){\circle*{1}}
\put(70,167){\circle*{1}}
\put(90,167){\circle*{1}}
\put(110,167){\circle*{1}}

\path(10,160)(110,160)
\path(10,140)(30,159)(90,159)(110,140)

\put(120,150){$\xCN  \xBc x \xfA abcd \xfA z \xBe $}
\put(4,150){(1)}

\path(10,110)(70,110)(90,90)(110,90)
\path(10,90)(30,109)(70,109)(90,89)(110,109)

\put(120,100){$  \xBc x \xfA abc \xfA z \xBe $}
\put(4,100){(2)}

\put(10,78){add paths equal on $abc$, different on $d$,
to compensate lacking paths in (1)}

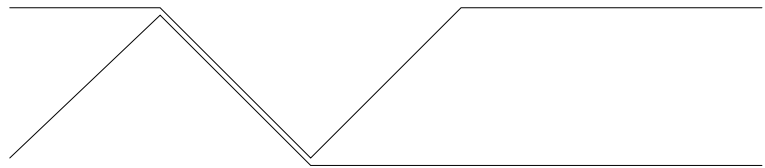
\path(10,60)(30,60)(50,40)(70,60)(110,60)
\path(10,40)(30,59)(50,39)(110,39)

\put(120,50){$\xCN  \xBc x \xfA ab \xfA z \xBe $}
\put(4,50){(3)}

\put(10,28){add paths different on $ab$, singletons on $c$,
so they don't disturb on $abc$:}
\put(10,24){seen on $abc$, the added paths are singletons, so
they respect automatically}
\put(10,20){$  \xBc x \xfA abc \xfA z \xBe $}

\end{picture}
}

\end{diagram}

\vspace{4mm}

$ \xCO $
\subsection{
Example of a rule derived from the basic rules
}

We will use the following definition.

\bd

$\hspace{0.01em}$


\label{Definition Func-Sigma-Mu}

Given $ \xbS $ as above, set

$ \xbS_{ \xbm }:=\{ \xBc X,Y,Z \xBe:$ $X,Y,Z$ are pairwise disjoint
subsets of $I,$ $ \xBc X \xfA Y \xfA Z \xBe \xce \xbS,$ but
for all $X' \xcb X$ and all $Z' \xcb Z$ $ \xBc X' \xfA Y \xfA Z \xBe \xbe
\xbS $ and $ \xBc X \xfA Y \xfA Z' \xBe \xbe \xbS \}.$

We will sometimes write $ \xBc X,X' \xfA Y \xfA Z \xBe $ etc. for $ \xBc X
\xcv X' \xfA Y \xfA Z \xBe.$

When we write $ \xBc X,X' \xfA Y \xfA Z \xBe $ etc., we will tacitly
assume that all sets
$X,X',Y,Z$ are pairwise disjoint.

\ed

\br

$\hspace{0.01em}$


\label{Remark Func-Sigma-Mu}

(1)
$ \xbS_{ \xbm }$ contain thus the minimal $X$ and $Z$ for fixed $Y,$ such
that $ \xBc X \xfA Y \xfA Z \xBe \xce \xbS.$

(2)
By rule (b), for all $ \xBc X \xfA Y \xfA Z \xBe \xbe \xbS,$ there is $
\xBc X',Y,Z' \xBe \xbe \xbS_{ \xbm }$ $X \xcc X',$ $Z \xcc Z',$
unless all $ \xbs,$ $ \xbt $ such that $ \xbs \xex Y= \xbt \xex Y$ can be
combined.

\er

As the cases can become a bit complicated, it is important to develop
a good intuition and representation of the problem. We do this now
in the proof of the following fact, where we use the result we want to
prove
to guide our intuition.

\bfa

$\hspace{0.01em}$


\label{Fact Func-Complic}

Let $ \xbS $ be closed under rules $(a)-(d).$
Then, if $ \xBc X,X',X'' \xfA Y \xfA Z,Z',Z'' \xBe \xbe \xbS_{ \xbm },$
then $ \xBc X,Z' \xfA X',Y,Z'' \xfA X'',Z \xBe \xce \xbS.$

\efa

\subparagraph{
Proof
}

$\hspace{0.01em}$


$ \xCO $

\vspace{10mm}

\begin{diagram}

\label{Diagram Func-Compos}

\centering
\setlength{\unitlength}{1mm}
{\renewcommand{\dashlinestretch}{30}
\begin{picture}(160,100)(0,40)

\path(10,120)(115,120)

\path(10,122)(10,118)
\path(25,122)(25,118)
\path(40,122)(40,118)
\path(55,122)(55,118)
\path(70,122)(70,118)
\path(85,122)(85,118)
\path(100,122)(100,118)
\path(115,122)(115,118)

\put(16,125){$X$}
\put(31,125){$X'$}
\put(46,125){$X''$}
\put(61,125){$Y$}
\put(76,125){$Z$}
\put(91,125){$Z'$}
\put(106,125){$Z''$}

\put(15,112){$\xbs_{X}$}
\put(30,112){$\xbs_{X'}$}
\put(45,112){$\xbs_{X''}$}
\put(56,112){$\xbs_{Y}=\xbt_{Y}$}
\put(75,112){$\xbt_{Z}$}
\put(90,112){$\xbt_{Z'}$}
\put(105,112){$\xbt_{Z''}$}

\path(10,110)(115,110)

\path(25,100)(115,100)
\put(118,100){(1)}

\path(10,95)(25,95)
\path(40,95)(115,95)
\put(118,95){(2)}

\path(10,90)(40,90)
\path(55,90)(115,90)
\put(118,90){(3)}

\path(10,85)(70,85)
\path(85,85)(115,85)
\put(118,85){(4)}

\path(10,80)(85,80)
\path(100,80)(115,80)
\put(118,80){(5)}

\path(10,75)(100,75)
\put(118,75){(6)}

\path(25,65)(40,65)
\path(55,65)(70,65)
\path(100,65)(115,65)
\path(25,64)(40,64)
\path(55,64)(70,64)
\path(100,64)(115,64)

\put(30,57)
{Prerequisite:
$\xbs_{X'}=\xbt_{X'}$, $\xbs_{Y}=\xbt_{Y}$, $\xbs_{Z''}=\xbt_{Z''}$}

\end{picture}
}

\end{diagram}

\vspace{4mm}

$ \xCO $

The upper line is the final aim.
Line (1) expresses that we can combine all parts except $s_{X},$ by
$ \xBc X',X'' \xfA Y \xfA Z,Z',Z'' \xBe,$ which holds by $ \xBc X,X'
,X'' \xfA Y \xfA Z,Z',Z'' \xBe \xbe \xbS_{ \xbm },$
by similar arguments, we can combine as indicated in lines
$(2)-(6).$
We now assume $ \xBc X,Z' \xfA X',Y,Z'' \xfA X'',Z \xBe \xbe \xbS.$ So
we have to look at fragments,
which agree on $X',Y,Z''.$ This is, for instance, true for (1) and (3).

We turn this argument now into a formal proof:

Assume

(A) $ \xBc X,Z' \xfA X',Y,Z'' \xfA X'',Z \xBe \xbe \xbS,$ and

(B) $ \xBc X,X',X'' \xfA Y \xfA Z,Z',Z'' \xBe \xbe \xbS_{ \xbm }.$

(C) $ \xBc X,X' \xfA Y \xfA Z,Z',Z'' \xBe $ by (B), see line (3)

(D) $ \xBc X \xfA X',Y,Z',Z'' \xfA X'',Z \xBe $ by (A) and rule (c)

(E) $ \xBc X \xfA X',Y \xfA Z,Z',Z'' \xBe $ by (C) and rule (c)

(F) $ \xBc X \xfA X',Y \xfA Z',Z'' \xBe $ by (E) and (b)

(G) $ \xBc X \xfA X',Y \xfA X'',Z,Z',Z'' \xBe $ by (D) and (F) and (d)

(K) $ \xBc X \xfA X',X'',Y \xfA Z,Z',Z'' \xBe $ by (G) and (c)

(L) $ \xBc X',X'' \xfA Y \xfA Z,Z',Z'' \xBe $ by (B), see line (1)

(M) $ \xBc Z,Z',Z'' \xfA X',X'',Y \xfA X \xBe $ by (K) and (a)

(N) $ \xBc Z,Z',Z'' \xfA Y \xfA X',X'' \xBe $ by (L) and (a)

(O) $ \xBc Z,Z',Z'' \xfA Y \xfA X,X',X'' \xBe $ by (M) and (N) and (d)

(P) $ \xBc X,X',X'' \xfA Y \xfA Z,Z',Z'' \xBe $ by (O) and (a).

So we conclude $ \xBc X,X',X'' \xfA Y \xfA Z,Z',Z'' \xBe \xbe \xbS,$ a
contradiction.

Comment:

We first move $Z',Z'' $ to the right, and then $X',X'' $ to the left.

Moving $Z',Z'':$

We use $X'' $ (or $Z)$ on the right, which not be changed, therefore we
can use
line (3), resulting in

(C) $ \xBc X,X' \xfA Y \xfA Z,Z',Z'' \xBe,$ or, directly

$(C')$ $ \xBc X,X' \xfA Y \xfA Z',Z'' \xBe,$ again by $ \xbS_{ \xbm },$

which is modified to

(F) $ \xBc X \xfA X',Y \xfA Z',Z'' \xBe,$ so we have on the right $Z'
,Z'' $ which we want to move.

We put $Z' $ in the middle ($Z'' $ is there already) of (A), resulting in

(D) $ \xBc X \xfA X',Y,Z',Z'' \xfA X'',Z \xBe.$

Now we can apply (d) to (D) and (F), and have moved $Z',Z'' $ to the
right:

(G) $ \xBc X \xfA X',Y \xfA X'',Z,Z',Z'' \xBe.$

We still have to move $X' $ and $X'' $ to the left of (G), and do this in
an
analogous way.

$ \xcz $
\\[3ex]

Note that our results stays valid, if some of the $X',X'',Z',Z'' $ are
empty.

Aber resultat darf nicht links oder rechts $ \xCQ $ sein.

\bco

$\hspace{0.01em}$


\label{Corollary Func-Complic}

Let $ \xbS $ be closed under rules $(a)-(d).$
Then, if $ \xBc X,X',X'' \xfA Y,Y',Y'' \xfA Z,Z',Z'' \xBe \xbe \xbS_{
\xbm },$ then
$ \xBc X,Y',Z' \xfA X',Y,Z'' \xfA X'',Y'',Z \xBe \xce \xbS.$

Thus, if, for given $Y \xcv Y' \xcv Y'',$
$ \xBc X,X',X'' \xfA Y,Y',Y'' \xfA Z,Z',Z'' \xBe \xbe \xbS_{ \xbm },$
then for no distribution of
$X \xcv X' \xcv X'' \xcv Y \xcv Y' \xcv Y'' \xcv Z \xcv Z' \xcv Z'' $ such
that the outward elements are
non-empty,
$ \xBc X,Y',Z' \xfA X',Y,Z'' \xfA X'',Y'',Z \xBe \xbe \xbS.$

\eco

\subparagraph{
Proof
}

$\hspace{0.01em}$


Suppose $ \xBc X,Y',Z' \xfA X',Y,Z'' \xfA X'',Y'',Z \xBe \xbe \xbS.$
Then by rule (c)
$ \xBc X,Z' \xfA X',Y,Y',Y'',Z'' \xfA X'',Z \xBe \xbe \xbS.$ Set
$Y_{1}:=Y \xcv Y' \xcv Y''.$ Then
$ \xBc X,Z' \xfA X',Y_{1},Z'' \xfA X'',Z \xBe \xbe \xbS,$ and
$ \xBc X,X',X'' \xfA Y_{1} \xfA Z,Z',Z'' \xBe \xbe \xbS_{ \xbm },$
contradicting
Fact \ref{Fact Func-Complic} (page \pageref{Fact Func-Complic}).
$ \xcz $
\\[3ex]
\clearpage
\section{
Examples of new rules
}
\subsection{
New rules
}

Above rules $(a)-(d)$ are not the only ones to hold, and we introduce now
more complicated ones, and show that they hold in our situation.
Of the possibly infinitary rules, only (Loop1) is given in full
generality, (Loop2) is only given to illustrate
that even the infinitary rule (Loop1) is not all there is.

For warming up, we consider the following short version of (Loop1):

\be

$\hspace{0.01em}$


\label{Example Loop1-Simple}

$ABC,ACD,ADE,AEB \xch ABE.$

We show that this rule holds in all $ \xbS.$

Suppose $A: \xbs,$ $B: \xbs = \xbt,$ $C: \xbt,$ so by $ \xCf ABC,$
there is $ \xbr_{1}$ such that

$A: \xbr_{1}= \xbs,$ $B: \xbr_{1}= \xbs = \xbt,$ $C: \xbr_{1}= \xbt.$
So by $ \xCf ACD,$ there is $ \xbr_{2}$ such that

$A: \xbr_{2}= \xbs,$ $C: \xbr_{2}= \xbr_{1}= \xbt,$ $D: \xbr_{2}= \xbt
.$ So by $ \xCf ADE,$ there is $ \xbr_{3}$ such that

$A: \xbr_{3}= \xbs,$ $D: \xbr_{3}= \xbr_{2}= \xbt,$ $E: \xbr_{3}= \xbt
.$ So by $ \xCf AEB,$ there is $ \xbr_{4}$ such that

$A: \xbr_{4}= \xbs,$ $E: \xbr_{4}= \xbr_{3}= \xbt,$ $B: \xbr_{4}= \xbt =
\xbs.$

So $ \xCf ABE.$

We abbreviate this reasoning by:

(1) $ \xCf ABC:$ $A: \xbs,$ $B: \xbs = \xbt,$ $C: \xbt $

(2) $ \xCf ACD:$ $(1)+ \xbt $

(3) $ \xCf ADE:$ $(2)+ \xbt $

(4) $ \xCf AEB:$ $(3)+ \xbt $

So $ \xCf ABE.$

It is helpful to draw a little diagram as in the following
Table \ref{Table Loop1-Short} (page \pageref{Table Loop1-Short}).

\ee

.
\begin{table}

\label{Table Loop1-Short}
\begin{center}
\begin{tabular}{|c|c|c|c|c|c|c|}
\hline
\multicolumn{7}{|c|}{Validity of $ ABC,ACD,ADE,AEB \xch ABE $} \\
\hline

 \xEH $ \xCf A$ \xEH $B$ \xEH $C$ \xEH $D$ \xEH $E$ \xEH \xEP
 \xEH $ \xbs $ \xEH $ \xbs = \xbt $ \xEH \xEH \xEH $ \xbt $ \xEH $ \xCf
ABE$? \xEP
\hline

(1) $ \xbr_{1}$ \xEH $ \xbs $ \xEH $ \xbs = \xbt $ \xEH $ \xbt $ \xEH \xEH
\xEH $ \xCf ABC$ \xEP
(2) $ \xbr_{2}$ \xEH $ \xbs $ \xEH \xEH $ \xbr_{1}= \xbt $ \xEH $ \xbt $
\xEH \xEH $ \xCf ACD$ \xEP
(3) $ \xbr_{3}$ \xEH $ \xbs $ \xEH \xEH \xEH $ \xbr_{2}= \xbt $ \xEH $
\xbt $ \xEH $ \xCf ADE$ \xEP
(4) $ \xbr_{4}$ \xEH $ \xbs $ \xEH $ \xbs = \xbt $ \xEH \xEH \xEH $
\xbr_{3}= \xbt $ \xEH $ \xCf AEB$ \xEP
\hline
\end{tabular}
\end{center}
\end{table}

We introduce now some new rules.

\bd

$\hspace{0.01em}$


\label{Definition New-Rules}

 \xEI

 \xDH
(Bin1)

$XYZ,XY' Z,Y(XZ)Y' \xch X(YY')Z$

 \xDH
(Bin2)

$XYZ,XZY',Y(XZ)Y' \xch X(YY')Z$

 \xDH
(Loop1)

$AB_{1}B_{2}, \Xl,AB_{i-1}B_{i},AB_{i}B_{i+1},AB_{i+1}B_{i+2}, \Xl
,AB_{n-1}B_{n},AB_{n}B_{1} \xch AB_{1}B_{n}$
$ \xDB $
so we turn $AB_{n}B_{1}$ around to $AB_{1}B_{n}.$

When we have to be more precise, we will denote this condition
$(Loop1_{n})$ to fix the length.

 \xDH
(Loop2)

$ABC,ACD,DAE,DEF,FDG,FGH,HFB \xch HBF:$

 \xEJ

\ed

The complicated structure of these rules suggests already that the ternary
relations are not the right level of abstraction to speak about
construction
of functions from fragments. This is made formal by our main result below,
which shows that there is no finite characterization by such relations.
In other words, the main things happen behind the screen.

\bfa

$\hspace{0.01em}$


\label{Fact New-Valid}

The new rules are valid in our situation.

\efa

\subparagraph{
Proof
}

$\hspace{0.01em}$


 \xEI

 \xDH
(Bin1)

(1) $ \xCf XYZ:$ $X: \xbs,$ $Y: \xbs = \xbt,$ $Z: \xbt $

(2) $XY' Z:$ $X: \xbs,$ $Y': \xbs = \xbt,$ $Z: \xbt $

(3) $Y(XZ)Y':$ $(1)+(2)$

So $X(YY')Z.$

 \xDH
(Bin2)

Let $X: \xbs,$ $Y: \xbs = \xbt,$ $Y': \xbs = \xbt,$ $Z: \xbt $

(1) $ \xCf XYZ:$ $X: \xbs,$ $Y: \xbs = \xbt,$ $Z: \xbt $

(2) $ \xCf XZY':$ $(1)+ \xbt $

(3) $Y(XZ)Y':$ $(1)+(2)$

So $X(YY')Z.$

 \xDH
(Loop1)

(1) $AB_{1}B_{2}:$ $A: \xbs,$ $B_{1}: \xbs = \xbt,$ $B_{2}: \xbt $

(2) $AB_{2}B_{3}:$ $(1)+ \xbt $

 \Xl.

(i-1) $AB_{i-1}B_{i}:$ $(i-2)+ \xbt $

(i) $AB_{i}B_{i+1}:$ $(i-1)+ \xbt $

$(i+1)$ $AB_{i+1}B_{i+2}:$ $(i)+ \xbt $

 \Xl.

$(\xCf n-1)$ $AB_{n-1}B_{n}:$ $(n-2)+ \xbt $

$(\xCf n)$ $AB_{n}B_{1}:$ $(n-1)+ \xbt $

So $AB_{1}B_{n}.$
 \xDH
(Loop2)

Let

(1) $ \xCf ABC:$ $A: \xbs,$ $B: \xbs = \xbt,$ $C: \xbt $

(2) $ \xCf ACD:$ $1+ \xbt $

(3) $ \xCf DAE:$ $2+ \xbs $

(4) $ \xCf DEF:$ $3+ \xbs $

(5) $ \xCf FDG:$ $4+ \xbt $

(6) $ \xCf FGH:$ $5+ \xbt $

(7) $ \xCf HFB:$ $6+ \xbs $

So $ \xCf HBF$ by $B: \xbs = \xbt.$

 \xEJ

Note that we use here $B: \xbs = \xbt,$ $E: \xbs = \xbt,$ $H: \xbs =
\xbt,$ whereas the other tripels
are used for other functions.

$ \xcz $
\\[3ex]

Next we show that the full (Loop1) cannot be derived from the basic
rules $(a)-(d)$ and (Bin1), and shorter versions of (Loop1).
(This is also a consequence of the sequel, but we want to point it out
right away.)

\bfa

$\hspace{0.01em}$


\label{Fact Loop-Indep}

Let $n \xcg 1,$ then $(Loop1_{n})$ does not follow from the rules
$(a)-(d),$ $(\xCQ),$ (Bin1), and the shorter versions of (Loop1)

\efa

\subparagraph{
Proof
}

$\hspace{0.01em}$


Consider the following set of tripels $L \xcv L' $ over
$I:=\{a,b_{1}, \Xl,b_{n}\}:$

$L:=\{ab_{1}b_{2},$  \Xl, $ab_{i}b_{i+1},$  \Xl, $ab_{n-1}b_{n},$
$ab_{n}b_{1}\},$

$L':=\{ \xCQ AB:$ $A \xcs B= \xCQ,$ $A \xcv B \xcc I\},$

and close this set under symmetry (rule (a)).
Call the resulting set $ \xda.$

Note that, on the outside, we have $ \xCQ $ or singletons, inside
singletons or $ \xCQ.$
If the inside is $ \xCQ,$ one of the outside sets must also be $ \xCQ.$

When we look at $L,$ and define a relation $<$ by $x<y$ iff $axy \xbe L,$
we see
that the only $<$-loop is $b_{1}<b_{2}< \Xl <b_{n}<b_{1}.$

We show first that $ \xda $ is closed under rules $(a)-(d)$
(see Definition 
\ref{Definition Basic-Rules} (page 
\pageref{Definition Basic-Rules})).

(a) is trivial.

(b) If $W= \xCQ $ or $Z= \xCQ,$ this is trivial, if $W=Z,$ this is
trivial, too.

(c) If $Z \xcv W= \xCQ,$ this is trivial, if $Z \xcv W$ is a singleton,
so
$Z= \xCQ $ or $W= \xCQ $ or $Z=W.$ $Z= \xCQ $ or $W= \xCQ $ are
trivial, otherwise $Z=W$ contradicts disjointness.

(d) $Z= \xCQ $ is trivial, so is $W= \xCQ,$ otherwise $Z=W$ contradicts
disjointness.

(Bin1)
$X= \xCQ $ or $Z= \xCQ $ are trivial, otherwise $X=Z$ is excluded by
disjointness.
So we are in $L' $ for $Y(XZ)Y'.$ So $Y= \xCQ $ or $Y' = \xCQ $ and it is
trivial.

Obviously, $(Loop1_{n})$ does not hold.

We show now that all $(Loop1_{k}),$ $0 \xck k<n$ hold.

The cases $n=1,$ $n=2$ are trivial.

Consider the case $2<k<n.$

This has the form $AB_{1}B_{2},AB_{2}B_{3}, \Xl,AB_{k-1}B_{k},AB_{k}B_{1}
\xch AB_{1}B_{k}.$

If $A= \xCQ $ or $B_{k}= \xCQ,$ the condition holds.

So assume $A,B_{k} \xEd \xCQ.$ Thus, by above remark, descending to
$B_{k-1}$ etc., we see that
all $B_{i} \xEd \xCQ,$ $1 \xck i \xck k.$
Thus, all prerequisites are in $L.$
Moreover, $ \xCf A$ has to be $ \xCf a,$ which
is the only element occuring repeatedly on the outside. Consider now the
relation $<' $ defined by $U<' V$ iff $ \xCf AUV$ is among the
prerequisites.
We then have $B_{1}<' B_{2}<'  \Xl <' B_{k}<' B_{1},$ where all $B_{i}$
are some $b_{j},$
we see that the resulting $<' $-loop is too short, so the prerequisites
cannot hold, and we have a contradiction.

$ \xcz $
\\[3ex]
\section{
There is no finite characterization
}

We turn to our main result.
\subsection{
Discussion
}

Consider the following simple, short, loop for illustration:

$ABC,ACD,ADE,AEF,AFG,AGB \xch ABG$ - so we can turn $ \xCf AGB$ around to
$ \xCf ABG.$

Of course, this construction may be arbitrarily long.

The idea is now to make $ \xCf ABG$ false, and, to make it coherent, to
make one
of the interior conditions false, too, say $ \xCf ADE.$ We describe this
situation fully, i.e. enumerate all conditions which hold in such a
situation.
If we make now $ \xCf ADE$ true again, we know this is not valid, so any
(finite)
characterization must say ``NO'' to this. But as it is finite, it cannot
describe all the interior tripels of the type $ \xCf ADE$ in a
sufficiently long loop,
so we just change one of
them which it does not ``see'' to FALSE, and it must give the same answer
NO, so
this fails.

Basically, we cannot describe parts of the loop, as the $< \xfA \xfA
>$-language
is not rich enough to express it, we see only the final outcome.

The problem is to fully describe the situation.
\subsection{
Composition of layers
}

A very helpful fact is the following:

\bd

$\hspace{0.01em}$


\label{Definition Sigma-S-H}

Let $ \xbS_{j}$ be function sets over $I$ into some set $K,$ $j \xbe J.$

Let $ \xbS $ $:=$ $\{$ $f:I \xcp K^{J}:$ $f(i)=\{ \xBc f_{j}(i),j \xBe:j
\xbe J,f_{j} \xbe \xbS_{j}\}$ $\}.$

So any $f \xbe \xbS $ has the form
$f(i)= \xBc f_{1}(i),f_{2}(i), \Xl,f_{n}(i) \xBe,$ $f_{m} \xbe \xbS_{m}$
(we may assume $J$ to be finite).

Thus, given $f \xbe \xbS,$ $f_{m} \xbe \xbS_{m}$ is defined.

\ed

\bfa

$\hspace{0.01em}$


\label{Fact Sigma-S-H}

For the above $ \xbS $ $ \xBc A \xfA B \xfA C \xBe $ holds iff it holds for all
$
\xbS_{j}.$

Thus, we can destroy the $ \xBc A \xfA B \xfA C \xBe $ independently, and
collect
the results.

\efa

\subparagraph{
Proof
}

$\hspace{0.01em}$


The proof is trivial, and a direct consequence of the fact that
$f=f' $ iff for all components $f_{j}=f'_{j}.$

Suppose for some $ \xbS_{k},$ $k \xbe J,$ $ \xCN  \xBc A \xfA B \xfA C \xBe.$

So for this $k$ there are $f_{k},f'_{k} \xbe \xbS_{k}$ such that
$f_{k}(B)=f'_{k}(B),$ but
there is no $f''_{k} \xbe \xbS_{k}$ such that $f''_{k}(A)=f_{k}(A),$
$f''_{k}(B)=f_{k}(B)=f'_{k}(B),$ $f''_{k}(C)=f'_{k}(C)$
(or conversely).
Consider now some $h \xbe \xbS $ such that $h_{k}=f_{k},$ and $h' $ is
like $h,$ but $h'_{k}=f'_{k},$ so
also $h' \xbe \xbS.$
Then $h(B)=h' (B),$ but there is no $h'' \xbe \xbS $ such that
$h'' (A)=h(A),$ $h'' (B)=h(B)=h' (B),$ $h'' (C)=h' (C).$

Conversely, suppose $ \xBc A \xfA B \xfA C \xBe $ for all $ \xbS_{j}.$
Let $h,h' \xbe \xbS $ such that $h(B)=h' (B),$ so for all $j \xbe J$
$h_{j}(B)=h'_{j}(B),$ where
$h_{j} \xbe \xbS_{j},$ $h'_{j} \xbe \xbS_{j},$ so there are $h''_{j} \xbe
\xbS_{j}$ with
$h''_{j}(A)=h_{j}(A),$ $h''_{j}(B)=h_{j}(B)=h'_{j}(B),$
$h''_{j}(C)=h'_{j}(C)$ for all $j \xbe J.$
Thus, $h'' $ composed of the $h''_{j}$ is in $ \xbS,$ and $h'' (A)=h(A),$
$h'' (B)=h(B)=h' (B),$ $h'' (C)=h' (C).$

$ \xcz $
\\[3ex]
\subsection{
Systematic construction
}

Recall the general form of (Loop1) for singletons:

$ab_{1}b_{2}, \Xl,ab_{i-1}b_{i},ab_{i}b_{i+1},ab_{i+1}b_{i+2}, \Xl
,ab_{n-1}b_{n},ab_{n}b_{1} \xch ab_{1}b_{n}$

We will fully describe a model of above tripels, with the exception of
$ab_{1}b_{n}$
and $ab_{i}b_{i+1}$ which will be made to fail, and all other $ \xBc X
\xfA Y \xfA Z \xBe $
which are not in above list of tripels to preserve, will fail, too
(except for $X= \xCQ $ or $Z= \xCQ).$

Thus, the tripels to preserve are:

$P$ $:=$ $\{ab_{1}b_{2}, \Xl,ab_{i-1}b_{i},$ (BUT NOT $ab_{i}b_{i+1})$
$,ab_{i+1}b_{i+2}, \Xl,ab_{n-1}b_{n},ab_{n}b_{1}\}$

We use the following fact:

\bfa

$\hspace{0.01em}$


\label{Fact Const-All-2-H}

Let $X \xcc I,$ $card(X)>1,$ $ \xbS_{X}$ $:=$ $\{$ $ \xbs:I \xcp \{0,1\}$
: $card\{x \xbe X: \xbs (x)=0\}$ is even $\}$

Then $ \xCN ABC$ iff $A \xcs X \xEd \xCQ,$ $C \xcs X \xEd \xCQ,$ $X \xcc
A \xcv B \xcv C.$

\efa

\subparagraph{
Proof
}

$\hspace{0.01em}$


``$ \xci $'':

Suppose $A \xcs X \xEd \xCQ,$ $C \xcs X \xEd \xCQ,$ $X \xcc A \xcv B
\xcv C.$

Take $ \xbs $ such that $card\{x \xbe X: \xbs (x)=0\}$ is odd, then $ \xbs
\xce \xbS_{X}.$
As $X \xcC A \xcv B,$ there is $ \xbt \xbe \xbS_{X}$ such that $ \xbs \xex
A \xcv B= \xbt \xex A \xcv B.$
As $X \xcC B \xcv C,$ there is $ \xbr \xbe \xbS_{X}$ such that $ \xbr \xex
B \xcv C= \xbs \xex B \xcv C.$
Thus, $ \xbt \xex B= \xbr \xex B.$
If there were $ \xba \xbe \xbS_{X}$ such that $ \xba \xex A \xcv B= \xbt
\xex A \xcv B$ and $ \xba \xex B \xcv C= \xbr \xex B \xcv C,$ then
$ \xba \xex A \xcv B \xcv C= \xbs \xex A \xcv B \xcv C,$ contradiction

``$ \xch $'':

Suppose $A \xcs X= \xCQ $ or $C \xcs X= \xCQ,$ or $X \xcC A \xcv B \xcv
C.$ We show $ \xCf ABC.$

Case 1:
$C \xcs X= \xCQ.$ Let $ \xbs, \xbt \xbe \xbS_{X}$ such that $ \xbs \xex
B= \xbt \xex B.$
As $C \xcs X= \xCQ,$ we can continue $ \xbs \xex A \xcv B$ as we like.

Case 2, $A \xcs X= \xCQ,$ analogous.

Case 3:
$X \xcC A \xcv B \xcv C.$ But then there is no restriction in $A \xcv B
\xcv C.$

$ \xcz $
\\[3ex]

We will have to make $ab_{1}b_{n}$ false, but $ab_{n}b_{1}$ true. On the
other hand,
we will make $ab_{1}b_{3}$ false, but $ab_{3}b_{1}$ need not be preserved.

This leads to the following definition, which helps to put order into
the cases.

\bd

$\hspace{0.01em}$


\label{Definition dmin-H}

Suppose we have to destroy $ \xCf axy.$ Then

$dmin(axy)$ $:=$ $min\{d(\{a,x,y\},\{a,u,v\}):$ $ \xCf auv$ has to be
preserved $\}$ - $d$ the
counting Hamming distance.

\ed

Thus, $dmin(ab_{1}b_{n})=0$ (as $ab_{n}b_{1}$ has to be preserved),
$dmin(ab_{1}b_{3})=1$
(because $ab_{1}b_{2}$ has to be preserved, but not $ab_{3}b_{1}).$

We introduce the following order defined from the loop prerequisites to be
preserved.

\bd

$\hspace{0.01em}$


\label{Definition Loop-Order-H}

Order the elements by following the string of sequences to be preserved as
follows:

$b_{i+1} \xeb b_{i+2} \xeb  \Xl  \xeb b_{n-1} \xeb b_{n} \xeb b_{1} \xeb
b_{2} \xeb  \Xl  \xeb b_{i-1} \xeb b_{i}$

Note that the interruption at $ab_{i}b_{i+1}$ is crucial here - otherwise,
there
would be a cycle.

As usual, $ \xec $ will stand for $ \xeb $ or $=.$
\subsection{
The cases to consider
}

\ed

The elements to consider are: $a,b_{1}, \Xl,b_{n}.$

Recall that the tripels to preserve are:

$P$ $:=$ $\{ab_{1}b_{2}, \Xl,ab_{i-1}b_{i},$ (BUT NOT $ab_{i}b_{i+1})$
$,ab_{i+1}b_{i+2}, \Xl,ab_{n-1}b_{n},ab_{n}b_{1}\}$

The $ \xBc X \xfA Y \xfA Z \xBe $ to destroy are (except when $X= \xCQ $
or $Z= \xCQ):$

 \xEh

 \xDH
all $ \xBc X \xfA \xfA Z \xBe $

 \xDH
all $ \xBc X \xfA Y \xfA Z \xBe $ such that $X \xcv Y \xcv Z$ has $>3$
elements

 \xDH
all tripels which do not have $ \xCf a$ on the outside, e.g. $ \xCf bgc$

 \xDH
and the following tripels:

(the (0) will be explained below - for the moment, just ignore it)

$ab_{1}b_{3}, \Xl,ab_{1}b_{n-1},$ $ab_{1}b_{n}$ (0)

$ab_{2}b_{1}$ (0), $ab_{2}b_{4}, \Xl,ab_{2}b_{n}$

$ab_{3}b_{1},$ $ab_{3}b_{2}$ (0), $ab_{3}b_{5}, \Xl,ab_{3}b_{n}$

 \Xl.

$ab_{i}b_{1},$ $ab_{i}b_{2},, \Xl,$ ALSO $ab_{i}b_{i+1}, \Xl
,ab_{i}b_{n}$

 \Xl.

$ab_{n-2}b_{1},, \Xl,$ $ab_{n-2}b_{n-3}$ (0), $ab_{n-2}b_{n}$

$ab_{n-1}b_{1},, \Xl,$ $ab_{n-1}b_{n-2}$ (0),

$ab_{n}b_{1},$ $, \Xl,ab_{n}b_{n-1}$ (0)

 \xEj
\subsection{
Solution of the cases
}

We show how to destroy all tripels mentioned above, while preserving
all tripels in $P.$

 \xEh
 \xDH
all $ \xBc X \xfA Y \xfA Z \xBe $ where $X \xcv Y \xcv Z$ has $>3$
elements:

See Fact 
\ref{Fact Const-All-2-H} (page 
\pageref{Fact Const-All-2-H})  with the $X$ there with 4 elements,
for all such $X,Y,Z$
separately, so all tripels in $P$ are preserved.

 \xDH
all $ \xBc X \xfA Y \xfA Z \xBe $ with 1 element: -

 \xDH
all $ \xBc X \xfA \xfA Z \xBe:$

This can be
done by considering $ \xbS_{j}:=\{0_{c},1_{c}\}.$ Then, say for
$a,c,$ we have to examine the fragments 00 and 11, but there is no 10 or
01.
For $ \xBc a \xfA b \xfA c \xBe $ this is no problem, as we have only the two
000,
111, which
do not agree on $b.$

 \xDH
all $ \xBc X \xfA Y \xfA Z \xBe $ with 2 elements: eliminated by $ \xBc X
\xfA \xfA Z \xBe $

 \xDH
all $ \xBc X \xfA Y \xfA Z \xBe $ with 3 elements:

 \xEh
 \xDH
$ \xCf a$ is not on the outside
 \xEh
 \xDH
$ \xCf a$ is in the middle, we need
$ \xCN xay:$ Consider $ \xbS $ with 2 functions, $0_{c},$ and the second
defined by
$a=0,$ and all $u=1$ for $u \xEd a.$
Obviously, $ \xCN xay.$ Recall that all tripels to be preserved have $
\xCf a$ on the
outside, and some other element $x$ in the middle. Then the two functions
are
different on $x.$

 \xDH
$ \xCf a$ is not in $ \xCf xyz,$ we need $ \xCN xyz:$
Consider $ \xbS $ with 2 functions, $0_{c},$ and the second defined by
$a=y=0,$ all $u=1$
for $u \xEd a,$ $u \xEd y.$
As $ \xCf a$ is neither $x$ nor $z,$ $ \xCN xyz.$ If some $ \xCf uvw$ has
$ \xCf a$ on the outside, say
$u=a,$ then both functions are 000 or 0vw on this tripel, so $ \xCf uvw$
holds.

 \xEj

 \xDH

$ \xCf a$ is on the outside, we destroy $ \xCf ayz:$

 \xEh

 \xDH
Case $dmin(ayz)>0$:

Take as $ \xbS $ the set of all functions with values in $\{0,1\},$ but
eliminate
those with $a=y=z=0.$
Then $ \xCN ayz$ (we have $100,001,101,$ but not 000),
but for all $ \xCf auv$ with $d(\{a,y,z\},\{a,u,v\})>0$ $ \xCf auv$ has
all possible
combinations,
as all combinations for $ \xCf ay$ and $ \xCf az$ exist.

 \xDH
Case $dmin(ayz)=0.$

The elements with $dmin=0$ are:

$ab_{1}b_{n},$ $ab_{2}b_{1},$  \Xl, $ab_{i}b_{i-1},$ NOT
$ab_{i+1}b_{i},$ $ab_{i+2}b_{i+1},$  \Xl, $ab_{n-1}b_{n-2},$
$ab_{n}b_{n-1},$
they were marked with (0) above.

$ \xbS $ will again have 2 functions, the first is always $0_{c}.$

The second function: Always set $a=1.$

We see that the tripels with $dmin=0$ to be destroyed have the form $ \xCf
ayz,$ where
$z$ is the immediate $ \xeb $-predecessor of $y$ in
above order - see Definition 
\ref{Definition Loop-Order-H} (page 
\pageref{Definition Loop-Order-H}).
Conversely, those to be preserved (in $P)$ have the form $ \xCf azy,$
where again
$z$ is the immediate $ \xeb $-predecessor of $y.$

We set $z' =1$ for all $z' \xec z,$ and $y' =0$ for all $y' \xed y.$
Recall that
$z \xeb y,$ so we have the picture
$b_{i+1}=1, \Xl,z=1,y=0, \Xl,b_{i}=0.$

Then $ \xCN ayz,$ as we have the fragments 000, 101. But $ \xCf azy,$ as
we have the
fragments 000, 110.
Moreover, considering the successors of the sequence, we give the values
11, or
10, or 00. This results in the function fragments for $ \xCf auv$ as 111,
or 110, or
100. But the resulting fragment sets (together with $0_{c})$ are then:
$\{000,111\},$ $\{000,110\},$ $\{000,100\}.$ They all make $ \xCf auv$
true. Thus, all
tripels in $P$ are preserved.

 \xEj
 \xEj
 \xEj
\clearpage
\section{
Systematic construction of new rules
}

This section is an outline - not a formal proof - for constructing a
complete rule set for our scenario.

We give here a general way how to construct new rules of the type
ABC, DEF,  \Xl. $ \xch $ XYZ which are valid in our situation.
\subsection{
Consequences of a single tripel
}

Let $(XX' X'')Y(ZZ' Z'')$ be a tripel, then all consequences of this
single tripel
have the form $X(X' YZ')Z$ (up to symmetry).

Obviously, such $X(X' YZ')Z$ are consequences, using rules (b) and (c).

We now give counterexamples to other forms, to show that they are not
consequences in our setting.
We always assume that the outside is not $ \xCQ.$
We consider $A=B=C=\{0,1\},$ and subsets of $A \xCK B \xCK C.$

 \xEh
 \xDH
$Y$ decreases:

Consider $\{000,111\},$ then ABC, but not $A \xCQ C.$

 \xDH
$Z$ increases:

Consider $\{000,101\},$ then $A \xCQ B,$ but not $A \xCQ (BC).$

 \xDH
$X$ goes from left to right:

Consider $\{000,110\},$ then (AB)C, but not $A(BC)$

 \xDH
$Y$ increases by some arbitrary $W:$

Consider $\{000,101,110,011\},$ then $A \xCQ C,$ but not ABC.

 \xEj
\subsection{
Construction of function trees
}

We can construct new functions from two old functions using tripels ABC,
so,
in a more general way, we have a binary function construction tree, where
the old functions are the leaves, and the new function is the root.
The form of such a tree is obvious, the tripels used are either
directly given, or consequences of such tripels. In
Example \ref{Example R-3} (page \pageref{Example R-3}), for
instance, in the construction of $ \xbr_{2},$ we used ACD, but we could
also
have used e.g. $AC(DD'),$ for some $D'.$
\subsection{
Derivation trees
}

Not all such function construction trees are proof trees for a rule
$T_{1}, \Xl,T_{n} \xch T,$ where the $T_{i}$ and $T$ are tripels.

We have to look at the logical structure of the tripels to see what we
need.
In order to show $T=ABC,$ we assume given two arbitrary functions $ \xbs $
and $ \xbt,$
which agree on $B,$ and construct $ \xbr $ such that on A $ \xbr = \xbs,$
on $B$ $ \xbr = \xbs = \xbt $
(the latter, $ \xbs = \xbt $ by prerequisite), and on $C$ $ \xbr = \xbt.$
We will write this as
$A: \xbr = \xbs,$ $B: \xbr = \xbs = \xbt,$ $C: \xbr = \xbt.$

Thus, we have no functions at the beginning, except $ \xbs $ and $ \xbt,$
so all leaves
in a proof tree for $T_{1}, \Xl,T_{n} \xch T$ have to be $ \xbs $ or $
\xbt.$ Moreoever, all we know
about $ \xbs $ and $ \xbt $ is that they agree on $B.$ Thus, we can only
use some
$T_{i}' =A' B' C' $ on $ \xbs $ and $ \xbt $ if $B' \xcc B.$ Likewise, in
the interior of the tree,
we can only use $ \xbs \xex B= \xbt \xex B,$ and, of course, all
equalities which hold be
construction. E.g., in
Example 
\ref{Example R-3} (page 
\pageref{Example R-3}), in the construction of $ \xbr_{2},$ by
construction of $ \xbr_{1},$ $C: \xbr_{1}= \xbt,$ so we can use ACD to
construct $ \xbr_{2}$ from
$ \xbr_{1}$ and $ \xbt.$

At the root, we must have a function $ \xbr $ of the form
$A: \xbr = \xbs,$ $B: \xbr = \xbs = \xbt,$ $C: \xbr = \xbt.$
In Example \ref{Example R-3} (page \pageref{Example R-3}),
$ \xbr_{4},$ at the root, was constructed using AEB from $ \xbr_{3}$ and $
\xbt.$
But we do not interpret $ \xbr_{4}$ as AEB, but as ABE, which is possible,
as $A: \xbr_{4}= \xbs,$ $B: \xbr_{4}= \xbs = \xbt,$ $E: \xbr_{4}= \xbt
.$

Intermediate nodes can be read as an intermediate result $A' B' C' $ by
the same
criteria: They must be functions $ \xbr' $ such that
$A': \xbr' = \xbs,$ $B': \xbr' = \xbs = \xbt,$ $C': \xbr' = \xbt $
and all $B'' $ such that
$B'': \xbs = \xbt $ used up to this node must be subsets of $B',$ as $B'
: \xbs = \xbt $ is the
only hypothesis we then have.
\subsection{
Universal trees
}
\subsubsection{
A proof for XYZ
}

The following is a universal proof for XYZ:

It is a binary tree, whose leaves are all $f$ or $g.$

It uses as prerequisite only $Y:f=g$ (and equalities constructed on the
way).

It makes $f,g,$ and all other functions as different as possible.

For instance,
in Example \ref{Example R-3} (page \pageref{Example R-3}), where we show that
$ABC,ACD,ADE,AEB \xch ABE,$ let us assume all sets A, etc. are singletons,
we then
set:
$ \xbs =00000,$ $ \xbt =10111,$ $ \xbr_{1}=00122,$ $ \xbr_{2}=03113,$ $
\xbr_{3}=04411,$ $ \xbr_{4}=00551.$
So each new node has a new default value $(2,3,4,5$ here).

Then we have no chance equalities, but only those we constructed.
In particular, if we write the equalities with $ \xbs,$ $ \xbt $ for
every $ \xbr_{i}$ thus
constructed, we can read off the derived equalities. There are no others.

The root of the tree must be a function $h,$ which agrees on $X,Y$ with
$f,$
and on $Y,Z$ with $g.$

This is a universal proof tree, as it works for any other pair $f,g,$ and
any
other internally constructed $ \xbr_{i},$ too.
\subsubsection{
Requirements for a proof for XYZ
}

Suppose we have a proof for XYZ.

We cannot assume we have anything but $f,g$ to start with.

The proof must be a binary tree, as the proof will be constructive, and we
have
no other construction principles but the combination of 2 functions.

So it is a binary tree, with leaves $f,g.$

It must also work for $f,g$ maximally different, i.e. outside $Y,$ they
may be
different. It must also work for the internal functions $ \xbr_{i}$
maximally
different. So we can only assume that $f,g$ agree on $Y,$ and all other
equalities must be by construction. Thus, it must also work for the
universal choice as done above. Assume now we have constructed this way
$h$
such that $h=f$ on $X,Y,$ $h=g$ on $Y,Z.$
This cannot be by coincidence, but it has to be a new function,
constructed
by the tree.
\subsubsection{
Summary: proofs for XYZ
}

To show XYZ, construct all universal trees for XYZ:

Begin with $f,g$ which agree at most on $Y,$ make them different
everywhere else.

Make all internal nodes different from each other by enumerating them,
and giving their number as default values to all other arguments.

Check if the root can be seen as the construction of a $h$ s.t. $h=f$ on
XY, $h=g$
on YZ.

If so, we have a proof of XYZ.

All proofs of XYZ have this form, as they must work for the universal
tree.
\clearpage
\subsection{
Examples
}

$ \xCO $

\vspace{10mm}

\begin{diagram}

\label{Diagram Rule-Const-1}

\centering
\setlength{\unitlength}{1mm}
{\renewcommand{\dashlinestretch}{30}
\begin{picture}(160,190)(0,70)

\put(60,182){Example \ref{Example R-1}}

\put(8,182){$\xbs$}
\put(42,182){$\xbt$}
\path(10,180)(25,165)
\path(27,165)(42,180)
\put(25,162){$\xbr_1$}

\put(60,162){$\xbt$}
\path(28,160)(43,145)
\path(45,145)(60,160)
\put(43,142){$\xbr_2$}

\put(90,122){Examples \ref{Example R-2} and \ref{Example R-4}}

\put(8,122){$\xbs$}
\put(42,122){$\xbt$}
\path(10,120)(25,105)
\path(27,105)(42,120)
\put(25,102){$\xbr_1$}

\put(53,122){$\xbs$}
\put(87,122){$\xbt$}
\path(55,120)(70,105)
\path(72,105)(87,120)
\put(69,102){$\xbr_2$}

\path(28,100)(48,80)
\path(50,80)(70,100)
\put(48,77){$\xbr_3$}

\end{picture}
}

\end{diagram}

\vspace{4mm}

$ \xCO $

$ \xCO $

\vspace{10mm}

\begin{diagram}

\label{Diagram Rule-Const-2}

\centering
\setlength{\unitlength}{1mm}
{\renewcommand{\dashlinestretch}{30}
\begin{picture}(160,195)(0,85)

\put(30,187){Example \ref{Example R-3}}

\put(8,182){$\xbs$}
\put(42,182){$\xbt$}
\path(10,180)(25,165)
\path(27,165)(42,180)
\put(25,162){$\xbr_1$}
\put(50,162){$\xbr_1$, using ABC - $A:\xbs, B:\xbs=\xbt, C:\xbt$}

\put(43,162){$\xbt$}
\path(26,160)(26,145)
\path(28,145)(41,160)
\put(25,142){$\xbr_2$}
\put(50,142){$\xbr_2$, using ACD - $A:\xbr_1=\xbs, C:\xbr_1=\xbt, D:\xbt$}

\put(43,142){$\xbt$}
\path(26,140)(26,125)
\path(28,125)(41,140)
\put(25,122){$\xbr_3$}
\put(50,122){$\xbr_3$, using ADE - $A:\xbr_2=\xbs, D:\xbr_2=\xbt, E:\xbt$}

\put(43,122){$\xbt$}
\path(26,120)(26,105)
\path(28,105)(41,120)
\put(25,102){$\xbr_4$}
\put(50,102){$\xbr_4$, using AEB - $A:\xbr_3=\xbs, E:\xbr_3=\xbt, B:\xbs=\xbt$}
\put(50,95){Interpretation: $ABE$, common part $B:\xbs=\xbt$}

\end{picture}
}

\end{diagram}

\vspace{4mm}

$ \xCO $

$ \xCO $

\vspace{10mm}

\begin{diagram}

\label{Diagram Rule-Const-3}

\centering
\setlength{\unitlength}{0.5mm}
{\renewcommand{\dashlinestretch}{30}
\begin{picture}(320,380)(0,200)

\put(90,372){Example \ref{Example R-5}}

\put(8,362){$\xbs$}
\put(42,362){$\xbt$}
\path(10,360)(25,345)
\path(27,345)(42,360)
\put(24,341){$\xbr_1$}

\put(53,362){$\xbs$}
\put(87,362){$\xbt$}
\path(55,360)(70,345)
\path(72,345)(87,360)
\put(68,341){$\xbr_2$}

\path(28,339)(48,319)
\path(50,319)(70,339)
\put(47,315){$\xbr_3$}

\put(108,362){$\xbs$}
\put(142,362){$\xbt$}
\path(110,360)(125,345)
\path(127,345)(142,360)
\put(124,341){$\xbr_1$}

\put(153,362){$\xbs$}
\put(187,362){$\xbt$}
\path(155,360)(170,345)
\path(172,345)(187,360)
\put(168,341){$\xbr_2$}

\path(128,339)(148,319)
\path(150,319)(170,339)
\put(147,315){$\xbr_4$}

\path(51,312)(96,267)
\path(100,267)(147,312)
\put(97,261){$\xbr_5$}

\put(153,282){$\xbs$}
\put(187,282){$\xbt$}
\path(155,280)(170,265)
\path(172,265)(187,280)
\put(168,261){$\xbr_1$}

\path(103,258)(135,227)
\path(142,227)(166,258)
\put(136,222){$\xbr_6$}

\end{picture}
}

\end{diagram}

\vspace{4mm}

$ \xCO $
\clearpage

Explanation:

By ``prerequisite'' of $ \xbr_{i}$ we mean the set $X$ we used in the
construction,
where $X: \xbs = \xbt.$ For instance, in the construction of $ \xbr_{2}$
in
Example \ref{Example R-1} (page \pageref{Example R-1}),
we used only that $B \xcv C: \xbr_{1}= \xbt $ by the construction of $
\xbr_{1},$ no additional
use of some $ \xbs = \xbt $ was made.

By ``common part'' of $ \xbr_{i}$ we mean the set $X$ such that $X:
\xbr_{i}= \xbs = \xbt.$

\be

$\hspace{0.01em}$


\label{Example R-1}

(Contraction), ABC, $A(BC)D$ $ \xcp $ $AB(CD)$:

(See Diagram 
\ref{Diagram Rule-Const-1} (page 
\pageref{Diagram Rule-Const-1})  upper part.)

 \xEI
 \xDH
$ \xbr_{1}:$ $A: \xbs,$ $B: \xbs = \xbt,$ $C: \xbt $

generated by $ \xCf ABC$ from $ \xbs,$ $ \xbt $

prerequisite $B,$

common part: $B$

$ \xbr_{1}$ can be interpreted as the (trivial) derived tripel $ \xCf ABC$
 \xDH
$ \xbr_{2}:$ $A: \xbr_{1}= \xbs,$ $B: \xbr_{1}= \xbs = \xbt,$ $C:
\xbr_{1}= \xbt,$ $D: \xbt $

generated by $A(BC)D$ from $ \xbr_{1},$ $ \xbt $

prerequisite -,

common part: $B.$

$ \xbr_{2}$ can be interpreted as a derived tripel by $AB(CD).$

$ \xbr_{2}$ can also be interpreted as a derived tripel by $A(BC)D$ or
$A(BD)C.$
Note that these possibilities can be derived from $AB(CD)$ by rule (c),
Weak Union.
 \xEJ

\ee

\be

$\hspace{0.01em}$


\label{Example R-2}

(Bin1), XYZ, $XY' Z,$ $Y(XZ)Y' $ $ \xch $ $X(YY')Z$:

(See Diagram 
\ref{Diagram Rule-Const-1} (page 
\pageref{Diagram Rule-Const-1})  lower part.)

\ee

 \xEI
 \xDH
$ \xbr_{1}:$ $X: \xbs,$ $Y: \xbs = \xbt,$ $Z: \xbt $

generated by $ \xCf XYZ$ from $ \xbs,$ $ \xbt $

prerequisite $Y$

common part: $Y$
 \xDH
$ \xbr_{2}:$ $X: \xbs,$ $Y': \xbs = \xbt,$ $Z: \xbt $

generated by $XY' Z$ from $ \xbs,$ $ \xbt $

prerequisite $Y' $

common part: $Y' $
 \xDH
$ \xbr_{3}:$ $Y: \xbr_{1}= \xbs = \xbt,$ $X: \xbr_{1}= \xbr_{2}= \xbs,$
$Z: \xbr_{1}= \xbr_{2}= \xbt,$ $Y': \xbr_{2}= \xbs = \xbt $

generated by $Y(XZ)Y' $ from $ \xbr_{1},$ $ \xbr_{2}$

prerequisites -

common part: $ \xCf YY' $

$ \xbr_{3}$ can be interpreted as a derived tripel by $X(YY')Z.$
 \xEJ

\be

$\hspace{0.01em}$


\label{Example R-3}

(Loop1) ABC, ACD, ADE, AEB $ \xch $ ABE:

(See Diagram \ref{Diagram Rule-Const-2} (page \pageref{Diagram Rule-Const-2}).)

 \xEI
 \xDH
$ \xbr_{1}:$ $A: \xbs,$ $B: \xbs = \xbt,$ $C: \xbt $

generated by $ \xCf ABC$ from $ \xbs,$ $ \xbt $

prerequisite $B$

common part $B$
 \xDH
$ \xbr_{2}:$ $A: \xbr_{1}= \xbs,$ $C: \xbr_{1}= \xbt,$ $D: \xbt $

generated by $ \xCf ACD$ from $ \xbr_{1},$ $ \xbt $

prerequisite -

common part -

$ \xbr_{2}$ cannot be interpreted as a derived tripel, as there was a
prerequisite
used in its derivation (B), but the common part in $ \xbr_{2}$ is $ \xCQ
.$
 \xDH
$ \xbr_{3}$ similar to $ \xbr_{2}:$

$ \xbr_{3}:$ $A: \xbr_{2}= \xbs,$ $D: \xbr_{2}= \xbt,$ $E: \xbt $

generated by $ \xCf ADE$ from $ \xbr_{2},$ $ \xbt $

prerequisite -

common part -

$ \xbr_{3}$ cannot be interpreted as a derived tripel, as there was a
prerequisite
used in its derivation (B), but the common part in $ \xbr_{3}$ is $ \xCQ
.$
 \xDH
$ \xbr_{4}:$ $A: \xbr_{3}= \xbs,$ $E: \xbr_{3}= \xbt,$ $B: \xbs = \xbt $

generated by $ \xCf AEB$ from $ \xbr_{3},$ $ \xbt $

prerequisites -

common part $B$

$ \xbr_{4}$ can be interpreted as the common part $B$ contains all
prerequisites
used in its derivation. $ \xCf ABE$ is the only non-trivial derived
tripel.

Note that we could, e.g., also have replaced ACD by $AC' (DC''),$ where
$C=C' \xcv C'',$
using rule (c), Weak Union.
 \xEJ

\ee

\be

$\hspace{0.01em}$


\label{Example R-4}

$BA(CD),$ $DF(CE),$ $(AB)(CD)(EF)$ $ \xch $ $B(ADF)(CE)$:

(See Diagram 
\ref{Diagram Rule-Const-1} (page 
\pageref{Diagram Rule-Const-1})  lower part.)

This example shows that we may need an assumption in the interior of the
tree (in the construction of $ \xbr_{3},$ we use $D: \xbs = \xbt).$
 \xEI
 \xDH
$ \xbr_{1}:$ $A: \xbs = \xbt,$ $B: \xbs,$ $C: \xbt,$ $D: \xbt $

generated by $BA(CD)$ from $ \xbs,$ $ \xbt $

prerequisites $ \xCf A$

common part $ \xCf A$
 \xDH
$ \xbr_{2}:$ $C: \xbt,$ $D: \xbs,$ $E: \xbt,$ $F: \xbs = \xbt $

generated by $DF(CE)$ from $ \xbs,$ $ \xbt $

prerequisite $F$

common part $F$
 \xDH
$ \xbr_{3}:$ A: $ \xbr_{1}= \xbs = \xbt,$ $B: \xbr_{3}= \xbs,$ $C:
\xbr_{1}= \xbr_{2}= \xbt,$ $D: \xbr_{1}= \xbr_{2}= \xbs = \xbt,$ $E:
\xbr_{2}= \xbt,$ $F: \xbr_{2}= \xbs = \xbt $

generated by $(AB)(CD)(EF)$ from $ \xbr_{1},$ $ \xbr_{2}$

prerequisite $D$

common part $ \xCf ADF$

So $ \xbr_{3}$ can be seen as the derived tripel $B(ADF)(CE)$ (but NOT as
$(AB)(DF)(CE)$
etc., as $ \xCf DF$ does not contain $ \xCf ADF.$
 \xEJ

\ee

\be

$\hspace{0.01em}$


\label{Example R-5}

$(AA')BC,$ $AD(CD'),$ $(AB')C(C' D),$ $(A' B')C(C' D'),$ $(AD)(B' CC'
)(A' D'),$
$BC(ADD')$ $ \xch $ $A(BD)(CD')$:

(See Diagram \ref{Diagram Rule-Const-3} (page \pageref{Diagram Rule-Const-3}).)

This example shows that we may need an equality
(here $ \xba $ and $ \xbb $ in the construction of $ \xbr_{5})$ which is
not
related to $ \xbs $ and $ \xbt.$ Of course, we cannot use it as an
assumption, but we
know the equality by construction.

$ \xba $ and $ \xbb $ will not be known, they are fixed, unknown
fragments.
 \xEI
 \xDH
$ \xbr_{1}:$ $A: \xbs,$ $A': \xbs,$ $B: \xbs = \xbt,$ $B': \xba,$
$C: \xbt $

generated by $(AA')BC$ from $ \xbs,$ $ \xbt $

prerequisites $B$

common part $B$
 \xDH
$ \xbr_{2}:$ $A: \xbs,$ $C: \xbt,$ $C': \xbb,$ $D: \xbs = \xbt,$ $D'
: \xbt $

generated by $AD(CD')$ from $ \xbs $ and $ \xbt $

prerequisite $D$

common part $D$
 \xDH
$ \xbr_{3}:$ $A: \xbs,$ $B': \xba,$ $C: \xbt,$ $C': \xbb,$ $D: \xbs
= \xbt $

generated by $(AB')C(C' D)$ from $ \xbr_{1}$ and $ \xbr_{2}$

prerequisite -

common part $D$
 \xDH
$ \xbr_{4}:$ $A': \xbs,$ $B': \xba,$ $C: \xbt,$ $C': \xbb,$ $D':
\xbt $

Generated by $(A' B')C(C' D')$ from $ \xbr_{1}$ and $ \xbr_{2}$

prerequisites -

common part -
 \xDH
$ \xbr_{5}:$ $A: \xbs,$ $A': \xbs,$ $B': \xba,$ $C: \xbt,$ $C':
\xbb,$ $D: \xbt,$ $D': \xbt $

generated by $(AD)(B' CC')(A' D')$ from $ \xbr_{3}$ and $ \xbr_{4}$

prerequisites - (note that equality on $B' $ and $C' $ is by construction
of $ \xbr_{3}$ and $ \xbr_{4},$ and not by a prerequisite on $ \xbs $ and
$ \xbt)$

common part: $D$
 \xDH
$ \xbr_{6}:$ $A: \xbs,$ $B: \xbs = \xbt,$ $C: \xbt,$ $D: \xbs = \xbt,$
$D': \xbt $

generated by $BC(ADD')$ from $ \xbr_{1}$ and $ \xbr_{5}$

prerequisites -

common part: $ \xCf BD$

Thus, $ \xbr_{6}$ may be seen as derived tripel $A(BD)(CD')$
 \xEJ

\ee

$ \xCO $
\chapter{
Subideal cases
}

$ \xCO $
\section{
The problem and the outline of a solution
}

$ \xCO $

\vspace{10mm}

\begin{diagram}

\label{Diagram Subideal}

\centering
\setlength{\unitlength}{1mm}
{\renewcommand{\dashlinestretch}{30}
\begin{picture}(160,120)(0,30)

\put(10,100){\circle*{1}}
\put(50,100){\circle*{1}}
\put(30,80){\circle*{1}}
\put(30,50){\circle*{1}}

\path(10,100)(30,80)(50,100)
\path(30,80)(30,50)

\path(11.3,97.3)(10,100)(12.7,98.7)
\path(47.3,98.7)(50,100)(48.7,97.3)
\path(29,77.2)(30,80)(31,77.2)

\put(5,99){$A$}
\put(52,99){$A'$}
\put(25,79){$B$}
\put(25,49){$C$}

\put(80,100){\circle*{1}}
\put(120,100){\circle*{1}}
\put(100,80){\circle*{1}}
\put(100,50){\circle*{1}}

\path(80,100)(100,80)(120,100)
\path(100,80)(100,50)
\path(100,50)(80,100)

\path(81.3,97.3)(80,100)(82.7,98.7)
\path(117.3,98.7)(120,100)(118.7,97.3)
\path(99,77.2)(100,80)(101,77.2)
\path(80.1,97)(80,100)(82,97.8)
\path(88,75)(92,76)

\put(75,99){$A$}
\put(122,99){$A'$}
\put(95,79){$B$}
\put(95,49){$C$}

\end{picture}
}

\end{diagram}

\vspace{4mm}

$ \xCO $

One of the advantages of defeasible inheritance systems is the ability to
treat
subideal cases.

In the left hand diagram,
(see Diagram \ref{Diagram Subideal} (page \pageref{Diagram Subideal})),
$C$ inherits from $B$ $ \xCf A$ and $A'.$ This is the ideal case.
In the right hand diagram, $C$ does not have property $ \xCf A,$ the
direct link $C \xcP A$
prevents this, but it still inherits $A' $ from $B,$ this is the subideal
case.

When we interpret $ \xCf A$ by ``blond'', $A' $ by ``tall'', $B$ by ``Swede'',
$C$ by a subset
of ``Swedes'', which are not blond, we have the classical dark haired Swedes
problem. Even dark haired Swedes should be tall.
Preferential structures have a problem with this, as they do not say
anything
about subideal cases (where not all properties which hold in the
minimal models, are valid).

Inheritance systems are modular in the following sense: the conditions
which are inherited are clearly and separately spelled out, $ \xCf A$ and
$A' $ here.
In preferential structures, we have - in principle - one tight knot
of ideal cases, and no way to separate the different properties - without
additional machinery. It is this machinery we want to examine here.

In inheritance systems, in principle, all combinations are possible:
$A \xcu A' \xcu B \xcu C,$ $A \xcu A' \xcu B \xcu \xCN C,$  \Xl, $ \xCN A
\xcu \xCN A' \xcu \xCN B \xcu \xCN C.$ We might not mention all,
but there is no contradiction to add nodes and arrows to make them
visible. E.g., we can introduce $D$ to one of the diagrams, with the
arrows $D \xcp A,$ $D \xcP A',$ $D \xcp B,$ $D \xcP C,$ etc. So, the
nodes code implicitly
logically independent possibilities, and we use this idea for
preferential structures.

Suppose we have a language $p,q,r,$ and a preferential structure where
$True \xcn p \xcu q \xcu r.$ Intuitively, we want to ``decompose'' this into
3 rules:
prefer $p$ over $ \xCN p,$ $q$ over $ \xCN q,$ $r$ over $ \xCN r.$ Note
that we can describe $ \xbm (True)$
by $p \xcu q \xcu r,$ but also by the conjunction of 7 rules, excluding
all other
models one by one: $ \xCN (pq \xCN r),$ etc. But these rules are not
independent:
There are cases with $ \xCN (pq \xCN r) \xcu \xCN (p \xCN qr),$ but there
is no case
with $(pq \xCN r) \xcu (p \xCN qr).$

So, the solutions seems to be, roughly: Find the finest (this exists,
see Fact 3.4 in  \cite{GS09b})
independent factorization $f_{1}, \Xl,f_{n}$ describing $ \xbm (X),$ and
for $X' \xcc X$ with
$X' \xcs \xbm (X)= \xCQ $ (when $X' \xcs \xbm (X) \xEd \xCQ,$
preferential structures take
care of this),
apply as many of the $f_{i}$ to $X' $ as possible. The ``as many''
should probably be determined by the subset relation, and not by counting,
as it is not sure that we are prepared to compensate the failure of one
$f_{i}$
by the validity of another $f_{i' }.$

So, we ``know'' how to inherit properties to subideal cases in preferential
structures.

Another basic idea of inheritance systems is specificity: Conflicts are,
if possible, solved by specificity. Tweety the penguin inherits egg-laying
from birds, but not-flying from penguins, and not flying from birds,
as penguins are more specific than birds. We have to carry this over to
our approach to preferential structures. The general situation is as
follows: We have a set $X,$ and inherit from $Y_{1}, \Xl,Y_{n}$ factors
$f_{1,1}, \Xl,f_{1,m_{1}}, \Xl,f_{n,1}, \Xl,f_{1,m_{n}},$ where the
$f_{i,j}$ are the factors of $ \xbm (Y_{i}).$
The $Y_{i}$ are partially ordered, and it seems natural to do some
``merger''
of the $f_{i,j},$ respecting priority determined by specificity. It is
probably adequate to take an ``axiom based'' approach, taking a suitable
subset of the $f_{i,j},$ as formalisms coming up with some compromise
(e.g. determined by some distance between models) are not only different
from the inheritance formalism, but will probably give unexpected results.

The situation is more complicated than in inheritance, as the $f_{i,j}$
need not
be independent when considering different $i$'s. Some approach like the
following is probably reasonable:

(1) Consider the strongest $Y_{i},$ ordered by specificity, and their
$f_{i,j}.$

(2) Consider all $f_{i,j}$ for those $Y_{i}.$ Identify minimal
inconsistent sets of
those $f_{i,j},$ and erase all $f_{i,j}$ involved (this corresponds to
direct
scepticism in inheritance), until a consistent set of $f_{i,j}$ is
obtained.

(3) Consider the next strongest $Y_{i' },$ and add similar to step (2) new
$f_{i',j},$
while preserving the $f_{i,j}$ already chosen, and considering consistency
together with the $f_{i,j}$ already chosen.

(4) Etc., until all $Y_{i}$ are done with.
\section{
Comments
}

 \xEh

 \xDH
We have here essentially a multi-valued approach. Not only classical
validity as maximally strong, and preferential structure as next
strongest,
but, partially ordered by specificity, arbitrarily many levels of
strength.

 \xDH
Note that preferential structures take care automatically of specificity
for the ideal case, basically, as we can handle all sets independently.
Here, we have to add a formalism to handle specificity.

 \xDH
Higher preferential structures, see
 \cite{GS08f}, can code our approach, but it is not sure that
the coding would be natural.

 \xDH
Independence as discussed above might need to be refined. For instance,
we might consider independence inside $X$ when considering $ \xbm (X).$

 \xDH
We can also ask whether we should not perhaps consider independence
of $X- \xbm (X),$ instead of independence of $ \xbm (X).$ The following
example gives an
answer:

\be

$\hspace{0.01em}$


\label{Example Mu-X}

Consider the language $\{p,q\}.$

(1)
Let $ \xbm (pq, \xCN pq,p \xCN q, \xCN p \xCN q)=pq.$ We have two rules,
$p< \xCN p,$ $q< \xCN q,$ and apply both.

(2)
$ \xbm (pq, \xCN pq,p \xCN q, \xCN p \xCN q)=(pq, \xCN pq,p \xCN q),$ we
avoid $ \xCN p \xCN q.$
But we do $ \xCf not$ avoid $ \xCN p$ and $ \xCN q,$ the rule is to avoid
one of them.
This does not seem to be such a good rule. In particular, factorization
as above does not work, contrary to the symmetric case (1).

\ee

 \xDH
Note that we can see the factorization of $ \xbm (X)$ as an approximation
of
the ideal case $ \xbm (X)$ by a set of rules.

 \xEj

$ \xCO $
\chapter{
Coding graphs by multisets
}

$ \xCO $
\section{
Introduction
}

This is a short comment on  \cite{AGS09}.

We examine here the coding of graphs by sets and multisets.

In the following, we abbreviate a set of labels or elements like
$\{a,b,c\}$ by
$ \xCf abc,$ etc.
\section{
Even the case with simple (not multi) sets is quite complicated
}

We consider here graphs generated by a subset of some powerset
(with the natural ordering by inclusion), and show that we need a certain
number of atomic labels to represent them.

The examples show that it is probably quite difficult to come up with a
minimal
number of elements - let alone working with multisets.
Example 
\ref{Example 6-El-2} (page 
\pageref{Example 6-El-2})  shows how complicated things can become.
There is an interplay of chains up and down, and antichains involved.

Thus, $ \xfI $ am quite sceptical about a good solution to the problem.

Moreover, Example 
\ref{Example 7-El} (page 
\pageref{Example 7-El})  shows that an inductive
construction is
impossible.

\be

$\hspace{0.01em}$


\label{Example 6-El-2}

Consider the graph generated by the subset
$\{abcdef,abcd,abc,ab,a,bcef,cef,ef,e,f\}$ of $ \xdp (abcdef).$

(A label of the node corresponding to)
$ \xCf abcd$ has to have at least 4 elements, by $a \xeb ab \xeb abc \xeb
abcd,$
and $ \xCf e,$ $ \xCf f$ have to have at least 4 elements
less than the top node $ \xCf abcdef.$

All maximal antichains have 3 elements, e.g., $\{abcd,e,f\}.$
The longest chains have 5 elements.

But we cannot use only 5 elements for labels, as any antichain containing
a node with 4 elements can have size at most 2.
\section{
There is no inductive algorithm by the natural ordering for the
simple set case
}

\ee

\be

$\hspace{0.01em}$


\label{Example 7-El}

This example shows that, in general, an inductive procedure is impossible.

Recall that, for a given set of $n$ elements, the number of subsets of
size
$m<n$ is $ \frac{n!}{(n-m)!*m!}.$

Take now a structure consisting of one antichain with 20 elements, and
nothing
else. This can be represented with 6 elements and subsets of size 3, as
$ \frac{6!}{3!*3!}$ $=$ $20.$

Take a structure with 4 antichains, each of size 20, and one above the
other. Thus, the size of the representing sets will increase at least by 1
from the lowest antichain to the next, etc.

As we are not allowed to look ahead, we begin again with subsets of size 3
of a set of 6 atomic labels for
the lowest antichain. So the next antichain must consist of sets of at
least
4 elements, the next of 5, the final of 6. Thus, we need at least 8
elements
for representation, as $ \frac{8!}{6!*2!}=28,$ 7 elements will not do.

If, however, we had begun with subsets of size 2 of a 7 element set, by
$ \frac{7!}{5!*2!}=21,$
we could have in the top layer only 5 element subsets, and this
is again possible, so 7 elements will do. But we have to look at the whole
structure to see this.
\clearpage
\section{
The multiset case
}

\ee

We work with a set of atomic labels $L:=\{a\} \xcv B,$ where $ \xCf a$ may
occur several
times,
this will be written $a^{n}$ for $n$ times $ \xCf a,$ etc.

We have the following trivial fact:

\bfa

$\hspace{0.01em}$


\label{Fact Compare}

(1) Let $B' \xcc B,$ then $a^{n}B' $ and $a^{m}B' $ are comparabel, as $n
\xck m$ or $m \xck n.$

(2) Let $B' \xcb B'' \xcc B,$ then $a^{n}B' $ is not comparabel to
$a^{m}B'' $ iff $n>m.$

\efa

\bco

$\hspace{0.01em}$


\label{Corollary Antichain}

(1) To code an antichain of size $2^{n},$ we need $B$ of size at least
$n.$

(2) We can code an antichain of size $2^{n}$ with $B$ of size $n.$

\eco

\subparagraph{
Proof
}

$\hspace{0.01em}$


(1) Suppose $B$ is smaller, then $card(\xdp (B))<2^{n},$ so two elements
of the antichain
are coded by the same $B' \xcc B,$ contradicting
Fact \ref{Fact Compare} (page \pageref{Fact Compare}), (1).

(2) Let $(B')$ be the exponent of the (unique by
Fact \ref{Fact Compare} (page \pageref{Fact Compare}), (1)) $a^{(B')}B'.$
Code the elements of the antichain by $\{a^{(B')}B':B' \xcc B\},$ where
$B' \xcb B'' $ implies
$(B'')<(B').$ Then the codes are pairwise incomparable by
Fact \ref{Fact Compare} (page \pageref{Fact Compare}), (2).
Note that $(\xCQ)$ is the biggest exponent, and $ \xCf (B)$ the
smallest.
(The idea is that, if the $B$-part of two codes is comparabel, we make
the $a$-part comparabel in the other direction, so the whole codes are
incomparabel.)

$ \xcz $
\\[3ex]

\be

$\hspace{0.01em}$


\label{Example Not-Isomorph}

Consider the structure $X \xeb Y$ and an isolated $Z.$

Obviously, we need at least one $b.$ We may code this
by $a \xeb a^{2},$ $b,$ or, by $b \xeb ab,$ $a^{2},$ and we have two,
non-isomorphic, codings.

\ee

In the following, we will code a bottom antichain of size $2^{n}$ by
$\{a^{(B')}B':B' \xcc B\},$ where $B$ has $n$ elements.
\section{
There is no inductive algorithm by the natural ordering for the
multiset case
}

We now show that an upward inductive algorithm, using the natural
ordering, is impossible. For this, we discuss progressively more
complicated examples. The last one,
Example 
\ref{Example Pyramide} (page 
\pageref{Example Pyramide}), is perhaps the most interesting, as it
shows
that we have to consider an arbitrarily deep and wide substructure
(with non-trivial interior nodes), to see that a decision taken lower
down cannot be upheld.

\be

$\hspace{0.01em}$


\label{Example No-Alg}

Consider an antichain of 4 elements at the bottom, say $ \xCf A,B,C,D.$

We might code this with the labels $abc,a^{2}b,a^{2}c,a^{3}.$

In the next level, we have an antichain of 2 elements, say $ \xCf X,Y,$
and
they have both the same predecessors, say $ \xCf A,B.$

Suppose $ \xCf A$ was coded by $a^{3},$ $B$ by $a^{2}b.$ Then we can code
$X$ by $a^{5}b,$ $Y$ by $a^{4}bc,$
and need no new label.

Suppose now that $ \xCf A$ was coded by $ \xCf abc,$ $B$ by $a^{2}b.$ Then
$X$ and $Y$ must include
$a^{2}bc,$ we may for instance make $X$ $a^{4}bc,$ but now we have to
introduce a new
variable, say $ \xCf d,$ and make $Y$ $a^{3}bcd.$

\ee

So, we have to look ahead. But it can be much more complicated. Take again
above example. Suppose we have now two antichains, $X,Y,$ and $X',Y',$
one
is above $A,B,$ the other above $C,D.$ Which one will have the $ \xCf
abc?$ If $X,Y$ is
higher than $X',Y',$ then we might have needed already $ \xCf d$
elsewhere, so we
can use it without additional cost. But $X$ might also be higher than $X'
,$ $Y' $
higher than $Y.$ What shall we do?

\be

$\hspace{0.01em}$


\label{Example Singletons}

This example shows that even the initial step of coding 4 elements
with 3 labels, as done above, might not always work: again, we have to
look
ahead.

Consider again an antichain of 4 elements at the bottom, say $ \xCf
A,B,C,D.$ Again,
we might code this with the labels $abc,a^{2}b,a^{2}c,a^{3}.$

Suppose we have in the second layer one new point above each pair from
$ \xCf A,B,C,D.$ One of the bottom nodes will be coded by $ \xCf a^{n}bc,$
another by $a^{n' }b,$
another by $a^{n'' }c.$ Suppose $n' \xcg n''.$ Let $X$ be above the
bottom elements coded by
$a^{n' }b$ and $a^{n}bc.$ Then it will also be above the bottom element
coded by $a^{n'' }c.$
But this is not wanted.

Thus, in this situation, we need a new label, say $d,$ to code the element
coded by $a^{n}bc.$

(We could also put the second layer nodes $X$ on ``stilts'', so they will
have
arbitrary height, like $a^{n}b \xeb a^{n+1}b \xeb a^{n+2}b \xeb  \Xl  \xeb
X,$ etc., so we have to climb
up arbitrarily high to see the problem.)

\ee

\bfa

$\hspace{0.01em}$


\label{Fact Basic}

Consider a bottom antichain with elements $a^{(B')}B',$ where $B' \xcc
B.$

Fix now $D \xcc B,$ let $D':=B- \xCf D,$ and consider $X:=\{a^{(D' E)}D'
E:$ $E \xcc D\}.$
Then, of course, $(D')=(D' \xCQ)>(D' E)$ for all $E \xEd \xCQ.$

Let $X=X' \xcv X'',$ where $X',X'' $ are disjoint and have
the same cardinality, and introduce two new nodes, $B' $ and $B'',$ such
that
$B' \xee A' $ for all $A' \xbe X',$ $B'' \xee A'' $ for all $A'' \xbe X''
,$ but for
no $A'' \xbe X'' $ $B' \xee A'',$ and for no $A' \xbe X' $ $B'' \xee A'
.$

Suppose without loss of generality
$a^{(D')}D' \xbe X'.$ Then there is $b \xbe D$ such that for no $a^{(D'
E)}D' E \xbe X',$
$b \xbe E.$ (Otherwise, by maximality of $(D'),$ all $x \xbe X$ would be
below $A'.)$

On the other hand, for cardinality reasons, there cannot be two such $b
\xbe D.$

$ \xcz $
\\[3ex]

\efa

\be

$\hspace{0.01em}$


\label{Example Pyramide}

Using $n+1$ atomic labels, $L:=\{a,b_{0}, \Xl,b_{n-1}\},$ we can code a
bottom
antichain $X_{0,0}, \Xl,X_{0,2^{n}-1}$ as follows: Work in the binary
system.
Set $B:=\{b_{0}, \Xl,b_{n-1}\},$ and code $B' \xcc B$ by $c(B'):= \xbS
\{2^{i}:b_{i} \xbe B' \}.$
This gives a natural total order on $ \xdp (B),$ and we use the inverse of
this
order for the exponent of $ \xCf a.$ Thus, as it should be, $(\xCQ)=(0,
\Xl,0)$ is the
biggest exponent, and $(B)=(1, \Xl,1)$ the smallest exponent.

In more detail, code $X_{0,i}$ by $a^{(i)}i,$ where $ \xCf i$ is written
in binary, $ \xCf i$
coding as above a subset of $B.$
Thus, $X_{0,0}$ is coded by $a^{(\xCQ)} \xCQ,$ $X_{0,1}$ by $a^{(0 \Xl
1)}0 \Xl 1=a^{(b_{0})}b_{0},$
$X_{0,2}$ by $a^{(0 \Xl 10)}0 \Xl 10=a^{(b_{1})}b_{1},$ etc., up to
$X_{0,2^{n}-1}=a^{(b_{n-1} \Xl b_{0})}b_{n-1} \Xl b_{0}.$

Then create new nodes above the bottom level, etc., always grouping
successive
lower nodes together, as follows:

$X_{1,0}, \Xl,X_{1,2^{n-1}-1}$

$X_{1,i} \xee X_{0,i*2},X_{0,i*2+1}$

$X_{k,0}, \Xl,X_{k,2^{n-k}-1}$

$X_{k,i} \xee X_{k-1,i*2},X_{k-1,i*2+1}$

up to $k=n-1$ (included).

The labelling of the new nodes is made by taking the union of lower
labels. Our
ordering of the exponents shows that this is possible, with exactly the
relations as defined.
See Diagram 
\ref{Diagram Pyramide} (page 
\pageref{Diagram Pyramide})  for an example with $n=3.$

\ee

$ \xCO $

\vspace{10mm}

\begin{diagram}

\label{Diagram Pyramide}

\centering
\setlength{\unitlength}{1mm}
{\renewcommand{\dashlinestretch}{30}
\begin{picture}(160,190)(0,0)

\put(8,26){\circle*{1}}
\put(28,26){\circle*{1}}
\put(48,26){\circle*{1}}
\put(68,26){\circle*{1}}
\put(88,26){\circle*{1}}
\put(108,26){\circle*{1}}
\put(128,26){\circle*{1}}
\put(148,26){\circle*{1}}

\put(18,42){\circle*{1}}
\put(58,42){\circle*{1}}
\put(98,42){\circle*{1}}
\put(138,42){\circle*{1}}

\put(38,74){\circle*{1}}
\put(118,74){\circle*{1}}

\put(58,138){\circle*{1}}
\put(98,138){\circle*{1}}

\multiput(98,45)(0,5){19}{\line(0,1){2}}

\path(9,27)(17,41)
\path(27,27)(19,41)
\path(49,27)(57,41)
\path(67,27)(59,41)
\path(89,27)(97,41)
\path(107,27)(99,41)
\path(129,27)(137,41)
\path(147,27)(139,41)

\path(19,43)(37,73)
\path(57,43)(39,73)
\path(99,43)(117,73)
\path(137,43)(119,73)

\path(18.5,44)(57,137)
\path(97.5,44)(59,137)
\path(58.5,44)(97,137)
\path(137.5,44)(99,137)

\put(5,20){$a^{(\xCQ)}$}
\put(25,20){$a^{(b)}b$}
\put(45,20){$a^{(c)}c$}
\put(65,20){$a^{(bc)}bc$}
\put(85,20){$a^{(d)}d$}
\put(105,20){$a^{(bd)}bd$}
\put(125,20){$a^{(cd)}cd$}
\put(145,20){$a^{(bcd)}bcd$}

\put(165,20){$X_{0,i}$}

\put(5,3){$000$}
\put(25,3){$001$}
\put(45,3){$010$}
\put(65,3){$011$}
\put(85,3){$100$}
\put(105,3){$101$}
\put(125,3){$110$}
\put(145,3){$111$}

\put(5,13){$X_{0,0}$}
\put(25,13){$X_{0,1}$}
\put(45,13){$X_{0,2}$}
\put(65,13){$X_{0,3}$}
\put(85,13){$X_{0,4}$}
\put(105,13){$X_{0,5}$}
\put(125,13){$X_{0,6}$}
\put(145,13){$X_{0,7}$}

\put(20,41){$a^{(\xCQ)}b$}
\put(60,41){$a^{(c)}bc$}
\put(100,41){$a^{(d)}bd$}
\put(140,41){$a^{(cd)}bcd$}

\put(165,41){$X_{1,i}$}

\put(40,73){$a^{(\xCQ)}bc$}
\put(105,73){$a^{(d)}bcd$}

\put(165,73){$X_{2,i}$}

\put(54,140){$a^{(\xCQ)}bd$}
\put(94,140){$a^{(c)}bcd$}

\end{picture}
}

\end{diagram}

\vspace{4mm}

$ \xCO $

For instance, $ \xCf (d)$ is the highest exponent in the right half, but
all
exponents on the left half are bigger than $ \xCf (d).$ Thus, all nodes on
the
right half are below $a^{(d)}bcd,$ and none on the left is below
$a^{(d)}bcd.$
For $a^{(\xCQ)}bc,$ all nodes on the right half contain $d,$ so they are
not
below $a^{(\xCQ)}bc,$ etc.
We add now two additional nodes, $a^{(\xCQ)}bd,$ and $a^{(c)}bcd.$ The
latter will
have more nodes below it than intended - see the broken line in the
diagram.
Consider first the node labelled $a^{(\xCQ)}bd.$ The nodes
below $a^{(cd)}bcd$ contain $ \xCf c,$ so they are not concerned, the same
holds
for those below $a^{(c)}bc.$ But it is impossible to add the node
$a^{(c)}bcd:$ By $(c)>(d)>(bd),$ we see that $a^{(d)}d \xeb a^{(c)}bcd$
and
$a^{(bd)}bd \xeb a^{(c)}bcd,$ a contradiction.

This is no accident, it does not depend on the
specific choice and distribution of the base labels, as we show now.
``(labelled  \Xl)'' refers to the example for $n=3,$ described in
Diagram \ref{Diagram Pyramide} (page \pageref{Diagram Pyramide}).

Consider, for an arbitrary labelling, $X_{n-1,0}$ (labelled by $a^{(\xCQ
)}bc)$
and $X_{n-1,1}$ (labelled by $a^{(d)}bcd)$ (these are all which
are on level $n).$ One of them has to be above $a^{(\xCQ)},$
without loss of generality,
let this be $X_{n-1,0}.$
Note that $(\xCQ)$ has to be the strictly biggest exponent, otherwise we
have no
antichain. One of the atomic labels, say $b_{j}$ $(\xCf d$ in the
diagram)
does not occur in the labelling
of $X_{n-1,0},$ otherwise, all bottom nodes would be below $X_{n-1,0}.$
For cardinality reasons, all others have to occur in the labelling of
$X_{n-1,0},$
see Fact \ref{Fact Basic} (page \pageref{Fact Basic}).
Moreover,
$b_{j}$ occurs in all labels of the bottom nodes below $X_{n-1,1},$ and
all
combinations of the other $b_{k}$ occur below $X_{n-1,1}.$ In particular,
we have
$a^{(b_{j})}b_{j}$ and $a^{(b_{n-1} \Xl b_{0})}b_{n-1} \Xl b_{0}$ below
$X_{n-1,1},$ and, by the same reasoning,
$(b_{j})$ is the strictly biggest exponent below $X_{n-1,1}.$

We split now $X_{n-1,0}$ into $X_{n-2,0}$ (labelled $a^{(\xCQ)}b)$
and $X_{n-2,1}$ (labelled $a^{(c)}bc)$ and repeat the argument,
using again Fact \ref{Fact Basic} (page \pageref{Fact Basic}).

Suppose, without loss of generality,
$a^{(\xCQ)} \xCQ $ is below $X_{n-2,0},$
so there must be some $a^{(b_{j' })}b_{j' }$ (labelled $a^{(c)}c)$
below $X_{n-2,1}.$ As $a^{(b_{j' })}b_{j' }$ is not
below $X_{n-1,1},$ $(b_{j' })>(b_{j}).$ Split now $X_{n-1,1}$ into
$X_{n-2,2}$ (labelled $a^{(d)}bd)$
and $X_{n-2,3}$ (labelled $a^{(cd)}bcd),$ and
suppose without loss of generality
$a^{(b_{j})}b_{j}$ is below $X_{n-2,2}.$ Create a new node $X$ (labelled
$a^{(c)}bcd)$
above $X_{n-2,1}$
and $X_{n-2,3}.$ Then it is bigger than $a^{(b_{j' })}b_{j' },$ so its
label has the exponent
$(b_{j' }),$ but it is also above $a^{(b_{n-1} \Xl b_{0})}b_{n-1} \Xl
b_{0}$ (labelled $a^{(bcd)}bcd),$
so it is also above
$a^{(b_{j})}b_{j},$ a contradiction by $(b_{j' })>(b_{j}).$
But we detect this only at level $n-2,$ and
we have to look at arbitrarily big subsets of the construction (in width
and depth!) to find a contradiction. Thus, in a strong sense, a recursion
is impossible.

$ \xDB $

Note that we may modify above example, e.g., introduce a smallest node
with
label $ \xCQ,$ and then lift the whole construction by adding everywhere
a new
set of labels, so we can embed it into an arbitrary diagram. Thus, the
problem is not only with the base level.
\clearpage
\section{
Generalization
}

We identify the different situations or objects (cameras, etc.) with
propositional models, and the properties with propositional variables.
The models may be defined only partially.

To distinguish different models, we name them. Thus, we might have
different
models with the same properties, but with different names. We assume that
all
values can only be 0/1 (the bull example needs more values).

$ \xfI $ think there are different ways to treat the situation:

 \xEh

 \xDH
We have only a local ranking, which is based on the values of the
propositional
variables. Based on this ranking, we try to complete the partially defined
models. Gaps are permitted (undefined values), if there is a gap, we just
forget this value for the ranking. If $m \xeb m' \xeb m'',$ and $m' (p)$
is undefined, then
we try to complete it, so that $m(p) \xck m' (p) \xck m'' (p).$

 \xDH
We have, in addition, a global ranking, where model $m$ may be considered
better
than model $m',$ for some external reason.

In this case, we try to complete the undefined values according to local
and
global ranking.

 \xDH
We have, in addition, a ranking of the propositional variables, where
$p$ might be stronger than $p',$ etc. In this case, we can work within
one
model, e.g., as follows: If $m(p)$ is ``positive'', and $m(p')$ unknown,
then
we assume that $m(p')$ is positive, too.

 \xEj

We then see the following:

 \xEh
 \xDH
We have a structure on the language, as 1 is better than 0.
In the third case above, we have an order on the variables, too, so even
more structure. See $p.$ 10 of our new book.
 \xDH
We may have a ``soft'' ranking, where some properties might be unknown,
then the known properties determine the ranking.

In this case, we fill in the unknown properties to coincide with the soft
ranking.
 \xDH
$ \xfI $ do not see why it is necessary to have only one (?) in the
matrix.

In particular, we may sometimes split 1 big matrix with two (?) into
2 small matrices with 1 (?) each.
 \xDH
This way of ordering reminds me of the ordering in deontic logic, where
situations may be better in several aspects.
 \xDH
It might be possible to generalize from elementary properties (propos.
variables) to formulas.
 \xDH
The locality of reasoning makes it likely that we have interpolation -
if we find a nice way to express it.
 \xDH
What are the laws of this reasoning? If we modify the matrices, what stays
constant, what changes, and how?
 \xDH
If we admit, say, 2 holes, we can examine Cumulativity:
Is the result the same, when we fill both at the same time, or, first 1,
then with the new matrix, 2?
 \xDH
$I$ think we can see this as a special case of preferential structures:
Replace the (?) with branching into 2 models, then prefer the one
which fits in better.
 \xDH
Vielleicht kann ich auch pref. Modelle wie oben als Matrix sehen, und dann
geometrisch arbeiten?
 \xDH
Mit Implikationen machen?
 \xDH
aus Bahnfahrt:
 \xEI
 \xDH
wieso nicht learning/detecting regularity?
 \xDH
hat an force bei $ \xCf a$ gedacht, nicht an min. labels, drum die vielen
Fehler
 \xDH
Ist das nicht detecting causality?
 \xDH
detect order, tendency
 \xDH
Ist Ansatz 0/1 einzusetzen, um zu sehen, was besser passt, gerechtfertigt?
Koennte das nicht eine Tendenz verschleiern?
 \xEJ

 \xEj

Dov,

$I$ have a few questions and remarks, which we might discuss on the phone:

(1) The problem differs from an interpolation problem, as, in the latter,
the
order is give, here it has to be found. Correct?

(2) Is finding regularities in one dimension (product, or model) really
the same
as finding them in the other dimension (properties)?

(3) $I$ am not sure that the coding of ``force'' by $ \xba^n$ is really what
you want,
and if the multiset approach is the right one. Do you have more on this?

(4) Detecting regularities is traditionally a learning problem, $I$ think.
Is
there a reason why this is not mentioned? Perhaps, we should work with
someone
from the learning community?

(5) You examine which of the possibilities give a better fit, 0 or 1 in
the
place of?. Does this always correspond to finding regularities? This
sounds
like a stupid question, but $I$ am not sure your answer is always true. If
so, it
might need a proof.

Karl

$ \xCO $
\chapter{
Re-considering some principles of non-monotonic logics
}

$ \xCO $
\section{
Introduction
}

We try to take a fresh look at some fundamental ideas of non-monotonic
logics.

In particular, we
 \xEh
 \xDH
examine the step from ``normally  \Xl'' to ``normal''
 \xDH
differentiate the consistency criterion of Reiter defaults
 \xDH
look at the ``inference greed'' of Reiter defaults, and other formalisms
like
inheritance, and give it an intuitive semantics through tentative theory
formation, and connect it to inductive reasoning
 \xDH
describe that specificity is not always a good criterion
 \xDH
suggest a more modular approach a la inheritance
 \xDH
examine subset systems more general than principal filters used in
preferential structures
 \xDH
describe how to generalize from propositional to first order defaults
 \xDH
introduce a notion of validity of a default in a classical model, and
describe
how to use it to solve conflicts and determine ``good'' models
 \xDH
finally, take a closer look at inheritance and motivate the use of
direct scepticism or of the intersection of extensions, and also
re-consider the translation of inheritance to other systems by examining
their language.
 \xEj

We stress those aspects which seem elementary, ``first principles''
to us, and try to
translate procedural aspects into a more declarative content.
The text is more questions and problems than answers.
\section{
General remarks
}
\subsection{
Not all defaults are about normality
}

Medical students are told: ``if you hear hoofbeat, think horses, not
zebras''.
The meaning is, of course, first think of normal, usual situations, and
not
exotic illnesses.
When we walk in the country, and hear the hissing of a snake, the advice
might
be: ``think rattle snakes, not garter snakes'', though the latter might be
more
common. The reason is, to treat first potentially dangerous situations.

Both describe default reasoning, but for different purposes (they can,
however,
both be summarized as ``useful'' reasoning, the first to treat common
situations,
the second to avoid dangers).
For the moment, we treat both as advice for acting (reasoning), or rules,
and
will write (hoofbeat:horse) and (hissing:rattler).
They are justified by different reasons, we have, so far, no formal
justification or semantics, and no way to treat a system of such rules.
But we are aware that the rules are ``rough'', it might be a zebra, it might
be a garter snake, after all.

Note that the default rule we chose to apply may depend on the context.
When we walk in the countryside, we use the cautious snake rule, when we
observe
from a safe position, we may use the rule that garter snakes are more
common
after all, so we conjecture it is a garter snake,
(hissing:garter-snake).
\subsection{
Systems of rules, subideal cases
}

We have many rules for birds, (birds:feathers), (birds:fly),
(birds:lay-eggs),
etc. When we write down all rules about birds, it might be that no single
bird satisfies all, the total set of rules for birds behaves like
the lottery paradox.
We may also have a mixture of rules with different motivations. In medical
diagnosis, one rule might be to check for a common and not so serious
illness,
another rule to exclude a rare, but dangerous and rapidly developping one.
We will probably decide about the latter first, then turn to the common
illness,
and if both are wrong, investigate further.
Note that we do not have here just ``normal'' and ``abnormal'' cases, but
three
classes - just as we sometimes have three cases to consider for a
mathematical proof.
\section{
Clarification of notions: Normality and consistency
}
\subsection{
Normality
}
\subsubsection{
From ``normally'' to ``normal''
}

There is an important - but often overlooked, see the author's own work -
change
from ``normally, birds fly'' to ``normal birds fly''.
The latter presupposes that normal birds, the ideal bird case, exist, the
former does not, it considers also partially normal birds. The ideal case
need
not exist, as the
lottery paradox shows. The intersection of the bird sets with ``normal''
properties might be empty - or meaninglessly small.
\subsubsection{
The behaviour of ``normal'' vs. finding normal elements
}

Preferential structures and their abstract treatment are about the
normal case. They investigate the properties of normality, of the ideal
case.
They do $ \xCf not$ investigate the subideal case, where only some
properties
of the ideal case are satisfied. This is done, implicitly, by Reiter
defaults, defeasible inheritance, etc., where we preserve as many normal
properties as possible.
Preferential structures
also do $ \xCf not$ investigate which elements (in the first order case)
are
normal, or as normal as possible. This is done by first order Reiter
defaults,
where as many elements as possible are made as normal as possible.
\subsection{
The consistency criterion for Reiter defaults (and other formalisms)
}

\label{Section Consistent}

A Reiter default is allowed to fire unless the consistency criterion is
violated.
But the inconsistency might be against a classical background theory, or
against
another default, or a combination of other defaults, etc.
In particular, criteria like specificity might be important.
Thus, a whole theory of elimination of inconsistencies may be necessary
to solve conflicts - as it is brought to light in defeasible inheritance.
In the first order case, which elements are normal, and to which degree,
is also
solved by an, implicitly, complicated theory.

Note that preferential structures have total control of minimal elements,
so there is no room for downward inheriting properties - unless we want to
work with special structures - and potential conflicts are obvious.
\section{
The implicit extension of conjectures
}
\subsection{
Inference greed
}

Reiter defaults (and, e.g., inheritance networks) are ``inference greedy''
in the
following triple sense:
 \xEh
 \xDH
The default $(: \xbf)$ will ``fire'', even if we know already $ \xbq,$ $(:
\xbf)$ is
implicitly broken down to subsets - contrary to preferential structures,
where
we do not have this homogeneity.
 \xDH
In the default set $\{(: \xbf),$ $(: \xbf')\},$ if $(: \xbf)$ cannot
fire (as $ \xCN \xbf $ holds),
$(: \xbf')$ may still be able to fire - in preferential structures, we
know nothing
beyond classical logic about not totally normal, ideal, elements,
whereas defaults can also treat the subideal case.
 \xDH
Open defaults $(: \xbf (x))$ make as many elements as possible normal,
i.e.
satisfy $ \xbf (x).$
 \xEj
\subsubsection{
A justification
}

It seems difficult to find a semantics in the usual sense for this
behaviour. Why should the world ``feel'' a pressure for normality?
Why should there be a direction towards maximal possible normality in the
world?

The only idea the present author had was to give an (informal) semantics
of both the world and our theory building about the world.
My, certainly naive, idea is in the platonic tradition. We make a theory
about
the world, knowing that it is only an approximation, but try to extend it
as far as possible (until contradictions - to be elaborated, see above,
Section \ref{Section Consistent} (page \pageref{Section Consistent})).
The basic assumption is that the world is regular, and we can, in
principle,
describe it in simple terms, but our description will not be perfect.
It is an assumption about homogeneity of the world, and independence of
properties, unless proven otherwise.
(It is also an exploratory approach: we explore the world, and try to be
conservative, in the sense of simplicity. As such, it has much in common
with
inductive reasoning.)

Thus, we have a pragmatic view, make as many defaults hold as possible,
also for subsets, and for as many elements as possible in the first order
case.
We do not seek ``best'' knowledge, about absolutely normal cases, but, more
modestly, distinguish between levels of knowledge, probabilities, like
truth
values in inheritance networks.
This can then be formalized by a simple relation of ``better'' between
models and elements, forgetting the human element of extending knowledge.
\subsection{
Remarks on specificity
}

The specificity criterion for deciding conflicts is one of the basic
tenets of non-monotonic reasoning.
If Tweety is a penguin, we conclude that the more specific information,
that
penguins don't fly, will win over the more general information that
birds fly. If there is no conflict, we assume that subsets behave like
supersets - see above.

The specificity criterion is fine for classification, as we assume that
many properties will be inherited from super- to subclass, but not all.
Subclasses may have a somewhat modified ``building plan''.
But specificity is irrelevant for other properties - for example for
``destructive''
properties. We will $ \xCf not$ try to find out if dead penguins can still
walk,
once we understood that dead animals cannot walk. Something in the
``construction'' of the animal has gone wrong, and we do not assume normal
life functioning any more. Thus, we have to distinguish properties which
``feel'' specificity, and those which do not.
(Likewise, we will not investigate how the dead specimens of a newly
discovered bird behave - we know it already, it is a ``transverse''
property, and
no inductive reasoning is necessary.)

This distinction goes beyond classical logic, as we distinguish different
types of predivates (or propositional variables, in the propositional
case).

``Penguin'' is not a capacity like flying, but a complex of properties.
Similarly,
we do diagnosis, e.g., for an illness, with $ \xCf distinctive$
properties,
which serve as indicators.

Note that specificity can be seen as an approximation: a more specific set
$B$ is
a better approximation to $ \xCf A$ than less specific set $C:$ $A \xcc B
\xcc C$
But we do not really work with
specificity as a set-wise relation: Tweety, a kolibri,
a blackbird, is a small set, but it seems useless. We need ``well defined''
small
sets, like penguins, we need property-wise or class-wise
(like penguin) approximation.
\subsection{
Induction
}

The justification for the inference greedy behaviour of defaults
makes a connection to inductive logic plausible.
Inductive reasoning is also inference greedy, we try to push our
knowledge as far as possible. Of course, the reasoning goes upward,
towards
the more general case, and not downward to subsets. Still, one should
explore further if there are common points.
In particular:

 \xEh
 \xDH
Is induction only inverse to the downward extension of knowledge of
defaults,
or are there deeper differences?
 \xDH
Can we transfer results and rules from one domain to the other?
 \xDH
Can we define inductive reasoning by the generalization which is best
extended downward in default reasoning (or vice versa)? So one will be
a reflection of the other?
 \xDH
Can we learn from ``real'' science, how physicists, or researchers in life
sciences, determine if a theory is thought to be sufficiently
corroborated?
What does ``practical philosophy of science'' say? How do they exclude
``disturbing influences''?
What does this mean for default reasoning?
Can we reflect this to default reasoning?
 \xDH
Can the degree of inconsistency
of Section 
\ref{Section Valid} (page 
\pageref{Section Valid})  be generalized to induction?
 \xEj
\section{
Modularity
}

An attractive feature of inheritance systems is their modularity.
Modularity corresponds also to the description of information as
approximation. We have several ``aims'', building blocks of a description,
and put them together as well as possible, in a principled way, based
on a basically modular world itself.

If we take this idea seriously, we have not one big language and theory,
but small fragments of non-monotonic theories, and - non-monotonic -
operators
on those fragments, which combine them, similar to a revision of
non-monotonic logics by non-monotonic logics. (Combining different
languages
is, e.g., a multiplication of models, etc.)
\section{
Subset systems beyond principal filters
}

Preferential structures (in the minimal version) generate principal
filters on sets, $ \xdf (X):=\{A: \xbm (X) \xcc A \xcc X\},$ together with
coherence
properties between filters over different sets, $X,$ $X',$ etc.
They have an intuitive interpretation by the notion of size.
The minimal elements are the ideal cases, and everything non-minimal is
negligeable, or small.

The lottery paradox and the limit version of preferential structures
motivate
to consider more general filters, or even weak filters.

Default systems also generate subset systems. E.g., $\{(: \xbf),(: \xbq
)\}$ generate
the ``good'' subsets $\{m:m \xcm \xbf \},$ $\{m:m \xcm \xbq \},$ $\{m:m \xcm
\xbf \xcu \xbq \},$ and perhaps
$\{m:m \xcm \xbf \xco \xbq \}.$ Considering the default system $\{(: \xbf
\xco \xbq),(: \xbq)\}$ shows that
$\{\{m:m \xcm \xbf \xco \xbq \},\{m:m \xcm \xbq \}\}$ and $\{\{m:m \xcm
\xbq \}\}$ should not be considered equivalent.
In the latter, only $\{m:m \xcm \xbq \}$ is ``good'', in the former, also
$\{m:m \xcm \xbf \xco \xbq \}$ will
be considered good, though not as good as $\{m:m \xcm \xbq \}.$ This is
intuitive, as
the default $(: \xbq)$ might not be able to fire, but the default $(:
\xbf \xco \xbq)$ may -
the system $\{(: \xbf \xco \xbq),(: \xbq)\}$ is not equivalent to the
system $\{(: \xbq)\}.$
If we interpret $ \xbq $ as the ideal case, then both describe the
same ideal case, or limit, but not the same subideal cases.
(This is like contrary-to-duty conditionals.)

Let $ \xdn (X)$ denote such abstract systems.

The following questions arise about $ \xdn (X):$
 \xEh
 \xDH
What are reasonable closure properties of $ \xdn (X)?$
 \xEI
 \xDH
A first idea is to
proceed as for deontic logic: Take all model sets derived from single
defaults, and close under union and intersection.
 \xDH
If $ \xcS \xdn (X)= \xCQ,$ we should probably consider only non-empty
intersections.
 \xDH
Should $X \xbe \xdn (X)?$ Probably not.
 \xDH
Is a system like $\{A,X-A\}$ reasonable? Are systems with $ \xcV \xdn
(X)=X$ reasonable?
 \xDH
Can different closure properties code different intuitions?
 \xDH
In which cases does $ \xdn (X)$ describe an approximation of ideal cases?
 \xEJ
 \xDH
Can we compare two different $ \xdn (X),$ $ \xdn' (X),$ e.g., if
$ \xcA A \xbe \xdn (X) \xcE A' \xbe \xdn' (X). A' \xcc A,$ then $ \xdn'
(X)$ is at least as sharp as $ \xdn (X)$ is?
 \xDH
What are reasonable coherence conditions between $ \xdn (X)$ and $ \xdn
(X')$
 \xDH
Can we find an intuitive interpretation of such systems, as we can
interpret $ \xbm (X)$ by size?
 \xDH
Can we generate such systems locally by a relation, as we did for $ \xbm
(X)?$
By higher order, reactive, relations?
 \xDH
If $ \xdn (X)$ is generated by a probability (as for the lottery paradox),
are there
special laws, resulting from substitution and sums?

E.g.: if $\{x,y\} \xce \xdn (X),$ $\{x,y' \},\{x',y\} \xbe \xdn (X),$
then $\{x',y' \} \xbe \xdn (X)?$
 \xDH
Are there intuitive ways to combine $A \xbe \xdn (X)$ with $A' \xbe \xdn
(X')$ to
$A \xCK A' \xbe \xdn (X \xCK X'),$ etc.?
 \xEj

Given $ \xdn (X),$ we can compare $x,x' \xbe X:$

\bd

$\hspace{0.01em}$


\label{Definition Compare}

Define $U(x):=\{A \xbe \xdn (X):x \xbe A\},$ and $S(x):= \xcS U(x).$

Let $x \xeb x' $ iff $S(x) \xcc S(x')$ (alternatively:
$card(S(x))<card(S(x')))$

This generalizes the comparison in preferential relations, minimal
elements are not comparable among each other.

\ed

\br

$\hspace{0.01em}$


\label{Remark Compare}

(1)
This is a special case of a preferential relation, as minimal elements
stay
minimal, it is about subideal elements.

(2)
It is robust under weakenings like in $\{N(\xbf),N(\xbf \xco \xbq)\}.$

(3)
What are the properties of the resulting relation, coherence conditions?

(4)
Can we find a complete set of such properties (representation)?

(5)
Transitivity of defaults is treated correctly: for $(\xbf: \xbq),$ $(
\xbq: \xbr),$ the
best $ \xbf $-models satisfy $ \xbq,$ and the best $ \xbq $-models
satisfy $ \xbr,$ so the overall
best $ \xbf $-models satisfy $ \xbr.$ This is not surprising, as we
pushed defaults
into the order, where we work with the best $ \xCf possible$ elements, as
in
preferential structures.

\er

\br

$\hspace{0.01em}$


\label{Remark Dynamik}

A remark on reasoning dynamics:

The full system $P$ has no dynamics, because of Cumulativity.
In the lottery paradox, once we concluded that $n$ will now win, $n' $ has
become
more likely to win. But, it could also be otherwise. If we conclude that a
bird will probably fly, the flying birds might even be more likely to
have feathers than the not flying ones. Thus, drawing conclusions might
also
make further conclusions more secure.
In inheritance, upward chaining adds new conclusions, but they become less
certain, as longer paths of reasoning offer more possibilities of attack.

There does not seem to exist a fully general theory of the dynamics of
reasoning - but this might also be too general a problem.
\section{
From propositional to first order logic
}

\er

In propositional logic, every (complete) possibility exists exactly once.
In 1st order logic, a predicate $p(.)$ may have 0, 1, many elements,
likewise
combinations of predicates, like $p(.) \xcu \xCN q(.).$
The combinations of properties correspond to propositional models.
Here, we treat these combinations, as if they were classical models. Then,
we put as many elements into the ``good'' combinations, and compare all
models
as in the propositional case.
Thus, we try
to put as many penguins as possible into the non-flying set, and the
others into
the flying set.
So, given a fixed universe $U,$ we prefer those
structures where more elements are ``good''.

 \xEh
 \xDH
More precisely, as in the propositional case, all cases are possible,
like $b(x) \xcu f(x),$ $b(x) \xcu \xCN f(x),$ etc., but they need not have
the same
cardinality. E.g., $b(x) \xcu \xCN f(x)$ might have 3 elements, $b(x) \xcu
f(x)$ 1 element,
or, vice versa. We prefer the latter, as the ``better'' case $b(x) \xcu
f(x)$ has
more elements than the ``less good'' case $b(x) \xcu \xCN f(x).$

Again, this is still up to interpretation for the right preference
relation.
This preference relation should certainly satisfy:
If, in structure $S,$ every $x$ in the universe satisfies a default set
$X_{x}$ which
is at least as good as the default set $X'_{x}$ satisfied in structure $S'
,$ then
$S$ should be preferred to $S'.$
More complicated relations may be considered, e.g., taking into account
cardinalities, like: More $x$ in $S$ satisfy ``good'' default sets than in
$S',$
etc., see Section \ref{Section Valid} (page \pageref{Section Valid}).

 \xDH
Suppose we have birds, penguins, sparrows.
Penguins cannot be flying birds, but sparrows should be. Sparrows are not
penguins, so, flying sparrows are better than not-flying sparrows. Flying
sparrows satisfy both defaults (not being penguins,
see Section \ref{Section Valid} (page \pageref{Section Valid})), but not-flying
sparrows violate the ``fly'' default, and satisfy the $(penguin: \xCN fly)$
default,
so they are worse.
We choose sparrows so that
they fall into the normal birds set, or, more precisely, among the
most normal birds. Names should be treated as unary predicates,
interpreted
by as normal as possible elements.

 \xEj
\section{
Validity of defaults and the best models
}

\label{Section Valid}

Consider the propositional case, and a non-nested default $(\xbf: \xbq
),$
i.e., $ \xbf $ and $ \xbq $ are classical formulas.
We treat the default similarly to the classical implication $ \xbf \xcp
\xbq,$ and
define for a classical model $m:$

$m \xcm (\xbf: \xbq)$ iff $m \xcm \xCN \xbf $ or $m \xcm \xbf \xcu \xbq
.$

We refine this. In classical logic, validity is absolute, 0 or 1.
We differentiate the strength of validity for defaults:

$m \xcm (\xbf: \xbq)$ holds with strength 1 (the strength of $m \xcm
\xCN \xbf)$ if $m \xcm \xCN \xbf.$

$m \xcm (\xbf: \xbq)$ holds with strength $M(\xbf)$ if $m \xcm \xbf
\xcu \xbq.$

$m \xcm (\xbf: \xbq)$ fails with strength $M(\xbf)$ if $m \xcm \xbf
\xcu \xCN \xbq.$

The strength $M(\xbf)$ takes care of specificity - the smaller $M(\xbf
),$ the bigger
the strength, this gives a partial order on strength.

For a full picture, we have to extend this definition to nested defaults.

\be

$\hspace{0.01em}$


\label{Example Valid}

Consider birds, penguin, ravens. Birds (including ravens) fly, penguins
don't,
penguins are birds, etc.
A penguin Tweety which does not fly, fails $ \xCf (bird:fly)$ with
strength ``bird'',
and satisfies $(penguin: \xCN fly)$ with strength ``penguin''. A penguin
Tweety' which
flies,
satisfies $ \xCf (bird:fly)$ with strength ``bird'', and fails $(penguin:
\xCN fly)$ with
strength ``penguin''. Tweety is a better model of the whole theory than
Tweety'
is, as Tweety fails for less strong defaults than Tweety' does.
Blacky, the flying raven, satisfies $ \xCf (birds:fly)$ with strength
``bird'', and
$(penguin: \xCN fly)$ with strength 1, as it is no penguin. Thus, Blacky
is
the best model of the theory (among Tweety, Tweety', Blacky).

\ee

We turn to the treatment of contradictions, this can be done in several
ways, defining a partial relation between models.
We outline requirements and possibilities, considering a theory
$T$ with classical information $ \xbf, \Xl $ and default information $(
\xbf: \xbq), \Xl $

 \xEh
 \xDH
Models which contradict classical information $ \xbf $ are the worst.
 \xDH
Models which contradict neither classical nor default information are
the best.
 \xDH
Fix a classical model $m.$ Let $S(m)$ be the (multi-) set of strengths
of defaults which $m$ fails. E.g., if $m \xcm \xbf \xcu \xbf',$ $m \xcm
\xCN \xbq \xcu \xCN \xbr \xcu \xCN \xbq',$
and $T$ consists of the defaults $(\xbf: \xbq),$ $(\xbf: \xbr),$ $(
\xbf': \xbq'),$ then
$S(m)=\{M(\xbf),M(\xbf),M(\xbf')\}.$ (We suppose that $ \xcM \xbq
\xcr \xbr $ - this has to be refined
to account for $ \xbq, \xbr $ which are not independent.)
 \xDH
A comparison of $m$ and $m' $ will be via a comparison of $S(m)$ with
$S(m').$

There are many possibilities:
 \xEh
 \xDH
We can treat $S(m)$ as a set, and forget multiple occurrences of the same
strength. This is probably unsatisfactory, as we will treat a model which
fails one default the same way as a model which fails many defaults - as
long
as they have the same strength. It results in usual preferential
structures,
which are unable to treat subideal cases.
$m \xcm \xbf \xcu \xbq \xcu \xCN \xbr $ will be then considered equivalent
to $m' \xcm \xbf \xcu \xCN \xbq \xcu \xCN \xbr $ (when
we consider just the defaults $(\xbf: \xbq),$ $(\xbf: \xbr)).$
 \xDH
We can consider $ \xcS S(m),$ and if $ \xcS S(m) \xcb \xcS S(m'),$
conclude that $m$ fails in
a worse way than $m' $ does.
 \xDH
We can combine (4.2) with a multiset approach, and ``count'' only if (4.2)
will
not decide between $m$ and $m'.$
 \xDH
We can use any other reasonable way to order a set of partially ordered
multisets.
 \xEj
The following questions arise:
 \xEh
 \xDH
These are special preferential relations, do additional properties hold?
 \xDH
Is there an abstract description, characterization, of such relations?
 \xEj
 \xDH
The first order case:

We use above partial order between propositional models to treat (open)
FOL defaults. A propositional model corresponds to a subset of the
universe,
like $X:=\{x:p(x) \xcu \xCN q(x)\}.$ If two such subsets $X,X' $ of the
universe
are comparable by above order, we prefer the model which has more
elements in the preferred $X$ - all other things being equal.

This is then a straightforward extension of the propositional case,
and handled in the same spirit.
 \xEj

\be

$\hspace{0.01em}$


\label{Example Contradict}

 \xEh
 \xDH
Consider the default set $ \xCf (p:q),$ $(p: \xCN q).$

Any $ \xCN p$-model satisfies both defaults, any $p$-model one, but not
the other.
So the globally best models are the $ \xCN p$-models, the best models of
$T=\{p,(p:q),(p: \xCN q)\}$ are all $p$-models.

 \xDH
This also gives an answer to the inconsistent default $(p: \xCN p):$
The globally best models are the $ \xCN p$-models, the best models for $p$
are all
$p$-models, being all equally bad.
 \xDH
Consider the default set $ \xCf (:p),$ $ \xCf (:q),$ $ \xCf (:r),$ and the
background theory
$(p \xcu q \xcu \xCN r) \xco (r \xcu \xCN p \xcu \xCN q).$ Then the model
$p \xcu q \xcu \xCN r$ satisfies 2 defaults,
the model $r \xcu \xCN p \xcu \xCN q$ only one. We decide by cardinality,
so the former
model is better.

(We need here that the defaults are ``decomposed'', e.g., not
$(:p \xcu q)$ instead of $ \xCf (:p),$ $ \xCf (:q).$ A finer treatment
might be needed to cover
cases like $(:p \xcu q).)$
 \xDH
We treat the Nixon diamond similarly. $ \xCf (p:r),$ $(q: \xCN r).$
Consider $T:=\{p \xcu q,(p:r),(q: \xCN r)\},$ the two models $p \xcu q
\xcu r$ and $p \xcu q \xcu \xCN r$ are
equally good (or bad), so none is preferred - we are directly sceptical,
we have no result about $r.$
 \xDH
We use specificity. For $ \xCf (b:f),$ $(p: \xCN f),$ $p \xcp b,$ we have
the globally best models:
$ \xCN b$-models and $b \xcu \xCN p \xcu f$-models (they satisfy both
defaults), among the
$p$-models,
(by $p \xcp b,$ $b$ holds) all fail one default, and the $ \xCN f$-models
are better
by specificity.
 \xDH
Consider $(: \xbf),$ $(: \xbf \xco \xbq).$ The best models are those
which satisfy $ \xbf,$ the
second best satisfy $ \xbf \xco \xbq $ but not $ \xbf,$ the worst satisfy
neither.
 \xEj
\subsection{
Asymmetric OR
}

\ee

We have treated $(\xbf: \xbq)$ above similarly to the classical
implication $ \xbf \xcp \xbq.$
It is natural to try and extend this, by translating ``somehow''
$(\xbf: \xbq)$ to $ \xbf \xcp (\xCN normal(\xbf) \xco \xbq).$ But
this is then an asymmetric ``OR'',
as in most cases, normality and $ \xbq $ will hold.
In particular, we will prefer to try and make normality hold, e.g., in
a mechanical proof system.

We may extend this idea to asymmetric theory revision, where $K*(\xbf
\xco \xbq)$
is preferably achieved by making $ \xbf $ true.
\section{
Inheritance
}

\br

$\hspace{0.01em}$


\label{Remark And}

Inheritance diagrams allow to treat subideal cases, but only with
information of differing strength; penguins still inherit ``feathers''
from birds, although they cannot fly.
Preclusion might override weaker
information. But we have an unrestricted AND for information of same
maximal strength (the ``Garbage In'' rule). So we cannot treat the Lottery
Paradox.
\subsection{
Direct scepticism vs. intersection of extensions
}

\er

We have to distinguish whether inheritance systems are to speak about the
state of the world, or about our knowledge of the world.
We may not know whether Nixon was a pacifist or not (direct scepticism),
but
he $ \xCf was$ one of the two, so one of the extensions represents reality
(leading
to the intersection of extensions approach).
Thus, the distinction between state of the world and knowledge,
and between elements and sets, provides an answer to the direct scepticism
vs.
intersection of extensions question.

In addition, if ``Nixon'' were a set, and not one element, there is even a
third
possibility: (almost) all Nixons are pacifists, (almost) all Nixons are
not
pacifists, and there is no majority for either. (In knowledge terms, we
may know
the latter holds, the latter or the first holds, $etc.,)$

\br

$\hspace{0.01em}$


\label{Remark Copies}

The existence of copies in classical preferential structures may code our
ignorance - we do not know which $x$ is smaller than $x',$ we only know
that
it is one of the $x \xbe X.$ We have all possibilities in one structure,
this
expresses scepticism.
Alternatively, we may work with many structures in parallel, see
 \cite{SGMRT00}, this corresponds to an extensions approach.
\subsection{
The language of inheritance
}

\er

 \xEh
 \xDH
The (implicit) language of inheritance is $ \xCf not$ sets and arrows, but
the atoms
are arrows, and the results are valid paths. Only in a latter step, valid
paths
are transformed into (soft) arrows. When $p \xcp q$ is an arrow in the
diagram,
neither $q \xcp p$ nor $q \xcP p$ need be in the language. Thus, a
comparison (soundness
and completeness) with the reasoning with
small sets etc. must $ \xCf only$ be about the information which can be
expressed
in the language of the diagram. Here, $q \xcP p$ may well be a result of
reasoning
with corresponding small sets, but we cannot compare it, as it is not in
the
language.
 \xDH
We see this (the language) also by the fact that we may have several paths
resulting in the same conclusion, but one might
be destroyed by further reasoning, and the other not.

 \xDH
We can define the language using admissible paths (concatenations of
arrows
pointing in the same directions, with at most one negative arrow, at the
end),
and/or their conclusions.
 \xEj

$ \xCO $

\end{document}